\documentclass[10pt]{article}
\usepackage{amssymb}
\usepackage{amsmath}
\usepackage{graphicx}
\usepackage{fancyhdr}
\usepackage{cite}
\usepackage{empheq}
\usepackage{dcolumn}
\usepackage{bm}
\usepackage{lipsum}
\usepackage{float}

\begin{document}

\newcommand{\bin}[2]{\left(\begin{array}{c}\!#1\!\\\!#2\!\end{array}\right)}

\huge

\begin{center}
D'yakov-Kontorovitch instability of shock waves in hot plasmas
\end{center}

\vspace{0.5cm}

\large

\begin{center}
Nadine Wetta, Jean-Christophe Pain\footnote{jean-christophe.pain@cea.fr} and Olivier Heuz\'e
\end{center}

\normalsize

\begin{center}
CEA, DAM, DIF, F-91297 Arpajon, France
\end{center}

\vspace{0.5cm}

\begin{abstract}
The D'yakov-Kontorovich stability criterion for spontaneous emission of acoustic waves behind shock fronts is investigated for high-temperature carbon, aluminum, silicon and niobium plasmas. The D'yakov and critical stability parameters are calculated along the principal Rankine-Hugoniot curve with an equation-of-state model in which the contribution of bound and free electrons is calculated through a relativistic quantum average-atom model, solving the Dirac equation. The pressure is determined using the stress-tensor formula in the relativistic framework. We find that the instability occurs at the end of the ionization of electronic shells, when the Hugoniot curve departs from the $\rho/\rho_0$=4 asymptote to tend to the $\rho/\rho_0$=7 limit. In such conditions, if the plasma is optically thick, the contribution of blackbody radiation to the EOS is dominant, and the system becomes always stable. Our results indicate that the conditions in which the instability takes place are different from previously published estimates, due to assumptions made in the corresponding equation-of-state models, especially as concerns the relativistic effects, and depend on the radiative opacity of the material.
\end{abstract}

\section{Introduction}

The large interest on the study of plasma instabilities is not only triggered by fundamental physics curiosity but stems from important plasma applications. Such instabilities are the result of any disturbance that may occur in a plasma parameter such as density, temperature, magnetic or electric field, or current. They dictate the plasma behavior and evolution \cite{KASELOURIS17}. Several theories predict that matter properties can lead to instabilities which can affect the well known steady shock wave propagation scheme, and cause major loss of efficiency in inertial confinement fusion (ICF) or misunderstanding of astrophysical phenomena \cite{PAIN07,KRITCHER14,ZAGHLOUL18}. Therefore, the experimental existence of these instabilities has to be investigated and their consequences mastered versus time and space. A first case of instability corresponds to the change of convexity of the equation of state \cite{JOUGUET04,JOUGUET04c,JOUGUET06,DUHEM09,BETHE42,ZELDOVICH46,THOMPSON71,THOMPSON73}. It can be discontinuous like phase transitions which lead to split shocks (laboratory impact experiments), or continuous (plasma phase transitions \cite{FORTOV07,FORTOV07b} found for instance in neutron star collisions) which generates a combination of shocks and isentropic compressions \cite{MENIKOFF89,HEUZE09b}. A second case is called the D'yakov-Kontorovich instability \cite{DYAKOV54,KONTOROVICH57}. Up to now no proof of its existence in the Universe has been found \cite{ALFEREZ17}. We study here the case of increasing pressure-volume relationship on the Hugoniot which is mainly obtained after the first ionization in plasmas. In a former study \cite{HEUZE09}, we explored the equation of state in this domain for three metals (aluminum, iron and copper) of very different densities, but found no possibility of its existence. Our recent calculations at much higher temperature and pressure, taking into account relativistic effects, showed us a possibility to reach these conditions. It is the purpose of the present work. 

The study of this instability in shock waves propagating in a medium has been a subject of research for many decades (see for instance \cite{BATES99,BATES00,BATES02,BATES12,BATES15}). It occurs from initial perturbations in the shock front, which may grow exponentially during the course of shock propagation, or arise from spontaneous emission of sound waves and entropy vortex waves \cite{MOND97} which take away energy from the shock wave, resulting in its decay. The first type may be reduced by minimizing initial perturbations in the system. The second kind, which has the form of a non-vanishing ``flutter'' on the surface of the shock wave without an increase or decrease of the perturbations, depends mainly on the material properties in particular thermodynamic conditions, i.e., on the equation of state (EOS). There are several experimental evidences of shock instability in ionizing and dissociating gases \cite{GLASS77,GRIFFITHS76}. It would be worth investigating whether experimental evidence of shock instability or unexplained events could be associated to some theoretically predicted phenomena.

The conditions for the onset of instability in a medium with an arbitrary EOS were first derived by D'yakov \cite{DYAKOV54} and Kontorovich \cite{KONTOROVICH57}. It is known that the EOS of the material affects the stability criteria to a great extent. D'yakov's analysis was again re-examined by Fowles and Swan \cite{FOWLES73,FOWLES76,FOWLES84,SWAN75}. Interaction of small perturbations of the form $e^{i(kx-\omega t)}$ with the planar shock front yields the stability criteria first derived by D'yakov and Kontorovich. It provides the conditions for which spontaneous emission of sound occurs in the shock compressed material. When a sinusoidal perturbing wave is incident on the shock front, from the compressed side, a reflected wave is produced. This wave is likely to carry shock energy away from the front. To hold momentum and energy conservation across the shock front, an entropy vortex wave, characterized by zero pressure perturbation, is also produced along with the reflected sound wave, whose amplitude depends on the reflection coefficient, which is itself dependent on the Rankine-Hugoniot curve and the frequency of the incident sound wave. Spontaneous sound emission occurs when the amplitude of this reflected wave is finite in the limit of vanishing amplitude of the incident wave. 

\vspace{5mm}

Let us consider the shocked state in the $(P,V)$ ($P$ being the pressure and $V$ the volume) diagram and its related slopes associated to different curves:

\begin{itemize}

\item $ \mathcal{H} $ the slope of the Hugoniot curve:

\begin{equation}
\mathcal{H}= - \left. \frac{dP}{dV} \right|_{\mathcal{H}},
\end{equation}

\noindent the subscript $\mathcal{H}$ meaning ``along the Hugoniot shock adiabat'',

\item $ \mathcal{R} $ the slope of the Rayleigh line:

\begin{equation}
\mathcal{R}=  \frac{P-P_0}{V_0-V} = j^2, 
\end{equation}

\noindent where $ j = \rho_0 D = \rho (D-u) $ is the mass flow ($\rho$ being the matter density), $D$ and $u$ being the shock and material velocities respectively, and 0 denotes the initial state (pole).

\item $ \mathcal{S} $ the slope of the isentropic curve:

\begin{equation}
\mathcal{S}= - \left. \frac{\partial P}{\partial V}\right|_S = \frac{c_s^2}{V^2},
\end{equation}

\end{itemize}

\noindent where $S$ represents the entropy and $c_s$ the sound speed of the compressed material. We deduce the Mach number

\begin{equation}
M^2 =\frac{ \mathcal{R} }{ \mathcal{S}  } = \left( \frac{ D - u }{ c_s  } \right)^2
\end{equation}

\noindent and introduce the D'yakov-Kontorovich parameter

\begin{equation}\label{dyako}
h = - \frac{ \mathcal{R} }{ \mathcal{H} }  = \jmath^2 \left. \frac{dV}{dP} \right|_{\mathcal{H}} = -(D-u)^2 \left. \frac{d \rho}{dP} \right|_{\mathcal{H}}.
\end{equation}

The D'yakov-Kontorovich instability occurs if

\begin{equation}
 \frac{ 1 - ( \zeta + 1) M^2}{ 1 + ( \zeta - 1) M^2 }  =  h_c < h < 1+2 M.
\end{equation}

\noindent where $\zeta=\rho/\rho_0$ denotes the compression of the material. The inequality $h < 1+2 M$ is always largely satisfied in the cases we consider. The quantity $h_c$, specified by Kontorovich \cite{KONTOROVICH57}, will be referred to as the ``critical parameter'' in the following. It is worth mentioning that, according to Landau and Lifshitz \cite{LANDAU}, the D'yakov-Kontorovich instability is a special form of instability, which is not an instability in the literal sense: the perturbation (ripples) created on the surface, continues indefinitely to emit waves without being damped or amplified. This is why some authors call the D'yakov-Kontorovich unstable shock fronts ``neutrally stable'', as opposed to the absolutely stable shock fronts (\emph{e.g.} in an ideal gas) in which all initial perturbations rapidly decay with time. 

\vspace{5mm}

To our knowledge, the D'yakov-Kontorovich instability has never been observed experimentally. Some numerical modelings have been performed \cite{BATES99,BATES00,BATES02,KONYUKHOV07}; for instance, Konyukhov \emph{et al.} \cite{KONYUKHOV09} carried out hydrodynamic simulations. In a previous paper Konyukhov \emph{et al.} \cite{KONYUKHOV04} had presented a numerical analysis of the nonlinear instability of shock waves for solid deuterium and for a model medium described by a properly constructed equation of state. In the work described in Ref. \cite{BATES12,BATES15}, the D'yakov-Kontorovich instability is supplanted by a corrugation instability for $h>h_c$ when a piston is included in the theory. This remedies the problem of multivalued solutions to projectile impact problems, as described by \cite{FOWLES73,FOWLES76,FOWLES84}.

The form of the electronic EOS used by Mond \emph{et al.} \cite{MOND97} does not incorporate the quantum details of electronic states (shell structure) and pressure ionization is accounted for by a Gaussian function. In Ref. \cite{DAS11}, Das \emph{et al.} consider an equation of state model relying on scaled binding energy for the cold contribution, mean-field theory for the ionic part, and a screened hydrogenic model with $\ell-$splitting for the contribution of bound electrons. In their modeling, the free electrons are treated within the non-relativistic semi-classical Thomas-Fermi model.

\vspace{5mm}

In the present work, we investigate D'yakov-Kontorovich instability with an EOS model relying on a full quantum self-consistent calculation of bound and free states solving Dirac equation through a relativistic quantum average-atom model. Our equation-of-state model is presented in Sec. \ref{sec2}. In Sec. \ref{sec3}, Hugoniot results are given for carbon (C, $Z$=6), aluminum (Al, $Z$=13), silicon (Si, $Z$=14) and niobium (Nb, $Z$=41) plasmas, and the shock stability is studied for the four elements in light of D'yakov-Kontorovich instability criterion. We find that the instability arises mostly at the end of the ionization of electronic shells, when the relativistic effects start to be important for the electrons. Our conclusions are compared to the ones of previous investigations (see Ref. \cite{DAS11}). In Sec. \ref{sec4}, we show that the accounting for the contribution of a blackbody radiation to the EOS tends to suppress the instability.

\section{Description of the equation-of state model}\label{sec2}

Throughout the paper, atomic units will be used, except that we keep $c$ instead of $1/\alpha$, where $c$ is the speed of light and $\alpha=e^2/\left(8\pi\epsilon_0a_0\right)$ in order to avoid confusion with the Dirac matrices $\vec{\alpha}$. In our model, the equation of state is built according to

\begin{equation}
\left\{
\begin{array}{l}
P(\rho,T)=P_c(\rho)+P_{\mathrm{th,i}}(\rho,T)+P_{\mathrm{th,e}}(\rho,T)\\
E(\rho,T)=E_c(\rho)+E_{\mathrm{th,i}}(\rho,T)+E_{\mathrm{th,e}}(\rho,T),
\end{array}
\right.
\end{equation}

\noindent where the quantities $E_c$ and $P_c$ (often referred to as the ``cold curve'') are respectively the pressure and internal energy at the temperature $T$=0 K and $P_{\mathrm{th,i}}$ and $P_{\mathrm{th,e}}$ represent respectively the thermal ionic and electronic pressures, and $E_{\mathrm{th,i}}$ and $E_{\mathrm{th,e}}$ the thermal ionic and electronic internal energies. A thermal quantity $Y_{\mathrm{th}}$ is defined as

\begin{equation}
Y_{\mathrm{th}}(T)=Y(T)-Y(T=0).
\end{equation}

\subsection{Cold curve and ionic contribution}

The so-called cold curve is provided by Thomas-Fermi calculations at $T$=0 K supplemented by the Barnes correction \cite{MORE88,BARNES67}. The ideal-gas approximation can be used for sufficiently high temperatures and densities. For dense and condensed media, however, this model is rather crude and it is important to take into account the interactions between ions, which belong to the so-called non-ideality effects. In order to take into account non-ideality corrections to the thermal motion of ions, we followed the work of Nikiforov \textit{et al}. \cite{NIKIFOROV87,NIKIFOROV05}, who used an approximation based on the calculation of the EOS of a One-Component Plasma (OCP) by the Monte Carlo method \cite{HANSEN73,POTEKHIN05,MASSACRIER11}. In the OCP model, it is assumed that ions with identical charge move in a homogeneous medium that carries a charge of opposite sign. The authors carried out molecular-dynamics simulations to calculate the distribution function of ions and their interaction energy, and derived interpolation formulas on the basis of these results. The ion contribution can be obtained using the Virial theorem; the ionic pressure and internal energy are given by

\begin{equation}\label{iocp}
\left\{
\begin{array}{l}
P_{\mathrm{i}}(\rho,T)=\rho k_BT+\frac{\rho}{3}\Delta E_{\mathrm{i}}(\rho,T)\\
E_{\mathrm{i}}(\rho,T)=\frac{3}{2}k_BT+\Delta E_{\mathrm{i}}(\rho,T),
\end{array}
\right.
\end{equation}

\noindent where the role of $\Delta E_{\mathrm{i}}$ is to account for non-ideality effects. In computations of the equation of state, it is convenient to modify the interpolation formulas of Hansen \cite{HANSEN73} for the interaction energy $\Delta E_{\mathrm{i}}$ given by the OCP model to ensure that it can still be applied for arbitrary temperatures and densities. The point is that in the OCP model, at values of the non-ideality parameter $\Gamma$ (which is also named ``coupling parameter'' and will be explicitely given later in Eq. (\ref{gam})) close to 158, a phase transition occurs. Since the physical accuracy of the OCP model is not beyond dispute, in practical computations, to simplify the calculation of the ion contribution to the EOS, one requires that at large values of the parameter $\Gamma$, the ion energy $E_{\mathrm{i}}$ will not exceed the asymptotic value $3k_BT$. We have, if the quantity $\Delta E_{\mathrm{i}}(\rho,T)$ is smaller than $3k_BT/2$, 

\begin{equation}\label{ig3}
\Delta E_{\mathrm{i}}(\rho,T)=k_BT\;\left[\Gamma^{3/2}\sum_{k=1}^4\frac{a_k}{(b_k+\Gamma)^{k/2}}-a_1\Gamma\right]
\end{equation}

\noindent where the coefficients $a_k$ and $b_k$, $k$=1, 4 are given in table \ref{tab1}, and $\Delta E_{\mathrm{i}}(\rho,T)=\Delta E_{\mathrm{i}}(T)=3k_BT/2$ otherwise.

\begin{table}
\begin{center}
\begin{tabular}{cccc}\hline\hline
$a_1$ & $a_2$ & $a_3$ & $a_4$\\\hline
-0.895929 & 0.11340656 & -0.90872827 & 0.11614773\\\hline\hline
$b_1$ & $b_2$ & $b_3$ & $b_4$\\\hline
4.666486 & 13.675411 & 1.8905603 & 1.0277554\\\hline\hline
\end{tabular}
\caption{Values of parameters $a_k$ and $b_k$, $k$=1, 4, involved in Eq. (\ref{ig3}).}\label{tab1}
\end{center}
\end{table}

\subsection{Electronic contribution to the EOS: quantum average-atom model}

In the present work, we use a relativistic quantum average-atom model \cite{PENICAUD08} following the work of Liberman \cite{LIBERMAN79} (see also Ref. \cite{WILSON06}). Such a model is often used to calculate the electronic contribution to the equation of state. It relies on a self-consistent computation of the electronic structure. In a spherically symmetric potential, the one-electron wavefunctions, solutions of Dirac equation, are of the form

\begin{equation}\label{psis}
\psi_s(\vec{r})\equiv\psi_{j\ell m}(\vec{r})=\left(\begin{array}{l}
\frac{1}{r}f(r)\Omega_{j\ell m}(\theta,\phi)\\
-\frac{i}{r}g(r)\Omega_{j\ell'm}(\theta,\phi)
\end{array}
\right),
\end{equation}

\noindent where $\Omega_{j\ell m}$ and $\Omega_{j\ell' m}$ are two spinors. $j$, $\ell$ and $m$ are quantum numbers associated respectively to the total angular momentum $J$, to the orbital angular momentum $L$ and its projection $L_z$. We define the quantum number $\ell'$ by: 

\begin{equation}
\ell'=\left\{\begin{array}{ll}
\ell+1\;\;\;\;\mathrm{if}\;\;\;\; j=\ell+1/2\\
\ell-1\;\;\;\;\mathrm{if}\;\;\;\; j=\ell-1/2.
\end{array}
\right.
\end{equation}

The radial functions $f$ and $g$ verify the equations

\begin{equation}
\left\{
\begin{array}{l}
\frac{df}{dr}=-\frac{\kappa}{r}f(r)-\frac{V_{\mathrm{eff}}(r)-c^2-\epsilon}{c}g(r)\\
\frac{dg}{dr}=\frac{V_{\mathrm{eff}}(r)+c^2-\epsilon}{c}f(r)+\frac{\kappa}{r}g(r)
\end{array}
\right.
\end{equation}

\noindent where

\begin{equation}
\left\{
\begin{array}{l}
\kappa=-(\ell+1)\;\;\;\;\mathrm{for}\;\;\;\; j=\ell+1/2,\\
\kappa=\ell\;\;\;\;\mathrm{for}\;\;\;\; j=\ell-1/2.
\end{array}
\right.
\end{equation}

The effective potential $V_{\mathrm{eff}}$ is assumed to be constant outside the cavity ($V_{\mathrm{eff}}(r)=V_{\infty}$ for $r\geq R$). Therefore, the solutions of Dirac equation are known for $r\geq R$, and the equation has to be solved for $r\leq R$. The inside and outside solutions are matched at $r=R$. 

The number of bound electrons reads

\begin{equation}
N_{\mathrm{bound}}=\sum_i X_iF(\epsilon_i,\mu),
\end{equation}

\noindent and the number of free electrons

\begin{equation}\label{nfree}
N_{\mathrm{free}}=\sum_{\kappa}\int_0^{\infty}X_{\kappa}(\epsilon)F(\epsilon,\mu)d\epsilon,
\end{equation}

\noindent where $F$ denotes the Fermi-Dirac distribution

\begin{equation}\label{fd}
F(\epsilon,\mu)=\frac{1}{e^{\beta(\epsilon-\mu)}+1},
\end{equation}

\noindent $\mu$ being the chemical potential and $\beta=1/\left(k_BT\right)$. The factor $X_i$ is

\begin{equation}
X_i=\int_{r\leq R}\psi_i^*(\vec{r})\psi_i(\vec{r})d^3r
\end{equation}

\noindent and one has

\begin{equation}
X_{\kappa}(\epsilon)=2|\kappa|\int_0^{R}\left[P_{\kappa}^2(\epsilon,r)+Q_{\kappa}^2(\epsilon,r)\right]r^2dr,
\end{equation}

\noindent where

\begin{equation}\label{relek}
\epsilon=c^2\sqrt{1+\frac{k^2}{c^2}}\;\;\;\;\mathrm{or}\;\;\;\;k=\sqrt{2\epsilon\left(1+\frac{\epsilon}{2c^2}\right)}.
\end{equation}

\noindent and $P_{\kappa}$ and $Q_{\kappa}$ represent the radial components of the free-electron spinor. 

The chemical potential is obtained from the electro-neutrality condition

\begin{equation}
N_{\mathrm{bound}}+N_{\mathrm{free}}=Z,
\end{equation}

\noindent $Z$ being the atomic number of the considered element. In the usual regime (typically $\mu/\left(k_BT\right)=\beta\mu$ less than 250), the Fermi distribution is far from the step function. In that case, Eq. (\ref{nfree}) can be split in two parts:

\begin{equation}
N_{\mathrm{free},0}=\sum_{\kappa}\int_0^{\epsilon_m}X_{\kappa}(\epsilon)f(\epsilon,\mu)d\epsilon
\end{equation}

\noindent and

\begin{equation}\label{nfree1}
N_{\mathrm{free},1}=\sum_{\kappa}\int_{\epsilon_m}^{\infty}X_{\kappa}(\epsilon)e^{-\beta(\epsilon-\mu)}d\epsilon
\end{equation}

\noindent with $\epsilon_m=\max(0,\mu+10/\beta)$. \noindent In Eq. (\ref{nfree1}), we can assume that $X_{\kappa}(\epsilon)$ can be approximated using the ideal wavefunctions of an electron gas: 

\begin{equation}
\left\{\begin{array}{l}
P_{\kappa}(\epsilon,r)=k\sqrt{\frac{k}{\pi\epsilon}}j_{\ell}(kr)\\
Q_{\kappa}(\epsilon,r)=k\sqrt{\frac{k}{\pi\epsilon}}s_{\kappa}\frac{\epsilon}{kc}j_{\ell'}(kr),
\end{array}\right.
\end{equation}

\noindent where $s_{\kappa}$ is the sign of κ and we get finally in that case

\begin{equation}
\sum_{\kappa}X_{\kappa}(\epsilon)=\frac{1+\epsilon/c^2}{\pi^2}k\left(\frac{4}{3}\pi R^3\right),
\end{equation}

\noindent with the maximal orbital quantum number $\ell_m\approx k_m R$, $k_m$ being related to $\epsilon_m$ through the second identity of Eq. (\ref{relek}).

The model imposes $f=g=0$ at $r=0$ and $r\rightarrow\infty$. Outside the cavity, the radial functions $f$ ang $g$ satisfying those boundary conditions are, for bound states, modified Bessel functions of the third kind \cite{ABRAMOWITZ65}, exponentially decreasing, and, for free states, combinations of Bessel functions of the first and second kinds, with decreasing amplitudes as $r\rightarrow\infty$. 

\begin{itemize}
\item Outside the cavity we have, for $\epsilon<V_{\infty}$ (bound states):
\end{itemize}

\begin{equation}
\left\{
\begin{array}{l}
f(r)=a_0c\frac{k}{V_{\infty}-\epsilon}rK_{\ell+1/2}(kr)\\
\\
g(r)=a_0rK_{\ell'+1/2}(kr),
\end{array}
\right.
\end{equation}

\noindent where $K_{n+1/2}$ ($n$ being an integer) are modified Bessel functions of the third kind and $a_0$ the normalization constant

\begin{equation}
a_0=\frac{1}{\int_0^{\infty}\left(f^2(r)+g^2(r)\right)^2dr}.
\end{equation}

\begin{itemize}
\item Outside the cavity we have, for $\epsilon>V_{\infty}$ (free states):
\end{itemize}

\begin{equation}
\left\{
\begin{array}{l}
f(r)=b_0c\frac{k}{\epsilon-V_{\infty}}r\left[\cos(\delta_{\ell})j_{\ell}(kr)-\sin(\delta_{\ell})n_{\ell}(kr)\right]\\
g(r)=b_0r\left[\cos(\delta_{\ell})j_{\ell'}(kr)-\sin(\delta_{\ell})n_{\ell'}(kr)\right],
\end{array}
\right.
\end{equation}

\noindent where the normalization factor $b_0$ and the wave number $k$ are 

\begin{equation}
\left\{\begin{array}{l}
b_0=\sqrt{\frac{2}{\pi}}\frac{k}{\sqrt{1+\frac{c^2k^2}{\left(\epsilon-V_{\infty}\right)^2}}}\nonumber\\
\nonumber\\
k=\sqrt{2\left(V_{\infty}-\epsilon\right)\left(1-\frac{\left(V_{\infty}-\epsilon\right)}{2c^2}\right)}.
\end{array}
\right.
\end{equation}

The matching of the solutions at the cavity radius provides the spectrum of bound energies and the phase shifts $\delta_{\ell}$.

The electronic structure (bound and free states) being known, the thermodynamic quantities can be calculated. We compute the internal energy as

\begin{equation}
E=K+U_{c,1}+U_{c,2}+E_{\mathrm{xc}}^{int},
\end{equation}

\noindent where the kinetic energy $K$ reads

\begin{equation}
K=\sum_bn_b(\epsilon_b-V_{\infty})X_b+\sum_{\kappa}\int_0^{\infty}n(\epsilon)\left(\epsilon-V_{\infty}\right)X_{\kappa}(\epsilon)d\epsilon-\int_{r\leq R}\rho(r)\left(V_{\mathrm{eff}}(r)-V_{\infty}\right)d^3r
\end{equation}

\noindent with

\begin{equation}
\rho(r)=\sum_{i=b,k}n_i\psi_i(r)^*\psi_i(r)
\end{equation}

\noindent and $n_i$ is the Fermi-Dirac occupation of state $i$ (see Eq. (\ref{fd})). We have, more precisely

\begin{equation}\label{fdp}
n_b=\frac{1}{e^{\beta\left(\epsilon_b-\mu\right)}+1}\;\;\;\;\mathrm{and}\;\;\;\;n(\epsilon)=\frac{1}{e^{\beta\left(\epsilon-\mu\right)}+1}.
\end{equation}

The effective potential is given by

\begin{eqnarray}\label{pottot}
\left\{
\begin{array}{l}
V_{\mathrm{eff}}(r)=V_c(r)+V_{\mathrm{xc}}(r)-\nu\;\;\;\;\mathrm{if}\;\;\;\;r\leq R,\\
V_{\mathrm{eff}}(r)=V_{\infty}\;\;\;\;\mathrm{if}\;\;\;\;r>R
\end{array}
\right.
\end{eqnarray} 

\noindent with

\begin{equation}
V_c(r)=-\frac{Z}{r}+\int_{r'\leq R}\frac{\rho(r')}{|\vec{r}-\vec{r'}|}d^3r',
\end{equation}

\begin{equation}
V_{\mathrm{xc}}(r)=\mu_{\mathrm{xc}}[\rho(r),T]
\end{equation}

\noindent and

\begin{equation}
V_{\infty}=\mu_{\mathrm{xc}}[\bar{\rho},T],
\end{equation}

\noindent where $\bar{\rho}$ is the density of the jellium. The exchange-correlation chemical potential is 

\begin{equation}
\mu_{\mathrm{xc}}[n,T]=\left.\frac{\partial}{\partial n}\left(nf_{\mathrm{xc}}[n,T]\right)\right|_T,
\end{equation}

\noindent where $f_{\mathrm{xc}}$ is the exchange-correlation free-energy density. 

\vspace{5mm}

Although the last decades have seen the emergence of increasingly sophisticated formulations for the exchange-correlation effects in the homogeneous electron gas, one finds very few finite-temperature exchange-correlation potentials in the literature (see for instance \cite{GUPTA80,PERROT84,GRIMALDI85,ICHIMARU87,KIYOKAWA07,KARASIEV14}). 

Some of the best functionals follow the production of acurate Quantum Monte Carlo data including many-body effects well beyond the so-called ``ring'' contribution of the earlier based RPA (Random Phase Approximation) works (see for instance Ref. \cite{GUPTA80}). Among them, Ichimaru \emph{et al.} derived formulas \cite{ICHIMARU87} based on RPIMC (Restricted Path Integral Monte Carlo) data, which encountered some success in Density Functional Theory. Recently, Karasiev \emph{et al.} proposed a new expression \cite{KARASIEV14} obtained in the same spirit. Their formulas reproduce efficiently the most recent Monte-Carlo data (available at the date of their work), with a global error of 0.5 \%, and a maximum one of 3.3 \%. Although we did not notice significant differences with Ichimaru's values, we preferred it for its practical use. Indeed, RPIMC provide discrete values of the internal energy data that were fitted using Pad\'e approximants. The resulting expression was then integrated in $r_s=\left[3/\left(4\pi n\right)\right]^{1/3}$ to obtain a formula for $f_{\mathrm{xc}}$ which presents a rather complicated dependence to the density $n$ of the electron gas, leading to a delicate derivation of the exchange-correlation potential $V_{\mathrm{xc}}$. Karasiev \emph{et al.} used both the internal energy and the kinetic energy RPIMC data to build a set of $f_{\mathrm{xc}}$ values on the same grid. The discrete $f_{\mathrm{xc}}$ were then fitted using the same Pad\'e approximants as Ichimaru \emph{et al.}. The resulting expression is easy to compute and simplifies the derivation of $V_{\mathrm{xc}}$. Both Ichimaru's and Karasiev's formulas respect the correct behaviour at $T$=0 (where the Kohn-Sham exchange contributions are recovered) and at very high $T$ (where $V_{\mathrm{xc}}$ and $f_{\mathrm{xc}}$ decay as $1/\sqrt{T}$ corresponding to the classical Debye-H\"uckel limit). Our exchange-correlation functional is non-relativistic. Indeed, since the relativistic effects are a correction to the exchange correlation their impact is expected to be very small as well. This is the reason why we believe that the non-relativistic exchange-correlation approximation is sufficient. Although we did not find any exchange-correlation functional including simulateneously finite-temperature and relativistic effects, relativistic effects have been studied in great detail at zero temperature, see for instance Ref. \cite{ELLIS77}. 

\vspace{5mm}

Finally, the parameter $\nu$ in Eq. (\ref{pottot}) is

\begin{equation}
\nu=f_{\mathrm{xc}}[\rho(R),T]-\mu_{\mathrm{xc}}[\bar{\rho},T]+\frac{\bar{\rho}}{\rho(R)}\left(\mu_{\mathrm{xc}}[\bar{\rho},T]-f_{\mathrm{xc}}[\bar{\rho},T]\right).
\end{equation}

The quantity $U_{c,1}$ represents the electron-nucleus Coulomb interaction energy 
 
\begin{equation}
U_{c,1}=-Z\int_{r\leq R}\frac{\rho(r)}{r}d^3r
\end{equation}

\noindent and $U_{c,2}$ the electron-electron Coulomb interaction energy 

\begin{equation}
U_{c,2}=\frac{1}{2}\int\int_{r,r'\leq R}\frac{\rho(r)\rho(r')}{|\vec{r}-\vec{r'}|}d^3rd^3r'.
\end{equation}

\noindent $E_{\mathrm{xc}}^{int}$ represents the exchange-correlation energy

\begin{equation}
E_{\mathrm{xc}}^{int}=\int_{r\leq R}\rho(r)\epsilon_{\mathrm{xc}}[\rho(r),T]d^3r,
\end{equation}

\noindent with

\begin{equation}
\epsilon_{\mathrm{\mathrm{xc}}}[n,T]=\left.\frac{\partial}{\partial\beta}\left(\beta f_{\mathrm{xc}}[n,T]\right)\right|_n.
\end{equation}

The total entropy reads

\begin{equation}
S=\sum_bX_b\left[n_b\ln n_b+(1-n_b)\ln(1-n_b)\right]+\sum_{\kappa}\int_0^{\infty}X_{\kappa}\left(\epsilon\right)\left[n(\epsilon)\ln n(\epsilon)+(1-n(\epsilon))\ln(1-n(\epsilon))\right]d\epsilon+S_{\mathrm{xc}},
\end{equation}

\noindent $S_{\mathrm{xc}}$ being the exchange-correlation entropy

\begin{equation}
S_{\mathrm{xc}}=\int_{r\leq R}\rho(r)s_{\mathrm{xc}}[\rho(r),T]d^3r,
\end{equation}

\noindent with

\begin{equation}
s_{\mathrm{xc}}[n,T]=-\left.\frac{\partial}{\partial T}f_{\mathrm{xc}}[n,T]\right|_n.
\end{equation}

Both $\epsilon_{\mathrm{xc}}$ and $s_{\mathrm{xc}}$ are easily derived from Karasiev's formula for $f_{\mathrm{xc}}$.

\subsection{Stress-tensor formula for the pressure}

In order to calculate the pressure due to bound and free electrons, we propose here to apply the stress-tensor formula in the relativistic formulation. Blenski and Ishikawa derived a formula in the semi-relativistic Pauli approximation \cite{BLENSKI95}. More recently, Maranganti and Sharma have published an elegant way to built the tensor, still in the non-relativistic case \cite{MARANGANTI10}. We propose to generalize their approach step by step to apply it to Dirac equation. Following Ref. \cite{MARANGANTI10}, let us introduce the density operator $\hat{\vec{A}}(\vec{r})$ by

\begin{equation}
\hat{\vec{A}}(\vec{r})=\frac{1}{2}\left\{\hat{\vec{A}},\delta(\vec{R}-\vec{r})\right\}=\frac{1}{2}\left(\hat{\vec{A}}\delta(\vec{R}-\vec{r})+\delta(\vec{R}-\vec{r})\hat{\vec{A}}\right),
\end{equation}

\noindent which satisfies

\begin{equation}
\frac{\partial\hat{\vec{A}}(\vec{r})}{\partial t}=\frac{1}{i}\left[\hat{\vec{A}}(\vec{r}),\hat{H}\right],
\end{equation}

\noindent $\hat{H}$ being the Hamiltonian of the system. After development of the commutator and anti-commutator, we get

 \begin{equation}
\frac{\partial\hat{\vec{A}}(\vec{r})}{\partial t}=\frac{i}{2}\left\{\hat{\vec{A}},\left[\hat{H},\delta(\vec{R}-\vec{r})\right]\right\}+\left\{\left[\hat{H},\hat{\vec{A}}\right],\delta(\vec{R}-\vec{r})\right\}
\end{equation}

\noindent and, if $\hat{\vec{A}}$ and $\hat{H}$ commute

\begin{equation}
\frac{\partial\hat{\vec{A}}(\vec{r})}{\partial t}=\frac{i}{2}\left\{\hat{\vec{A}},\left[\hat{H},\delta(\vec{R}-\vec{r})\right]\right\}
\end{equation}

At this stage, Maranganti and Sharma pursue with the Hamilton operator $\hat{H}$. We follow the same procedure for the Dirac operator; the commutator $\left[\hat{H},\delta\right]$ is replaced by

\begin{equation}
\left[c\vec{\alpha}.\vec{p},\delta(\vec{R}-\vec{r})\right]=\left[c\sum_{j=1}^3\alpha_jp_j,\delta(\vec{R}-\vec{r})\right]
\end{equation}

\noindent or

\begin{equation}\label{com}
\left[c\sum_{j=1}^3\alpha_j\left(\frac{\hbar}{i}\nabla_j\right),\delta(\vec{R}-\vec{r})\right].
\end{equation}

In the two latter equations, $\vec{\alpha}$ represents the Dirac matrices

\begin{equation}
\vec{\alpha}=
\begin{pmatrix}
0 & \vec{\sigma}\\
\vec{\sigma} & 0
\end{pmatrix}
\end{equation}

\noindent where $\vec{\sigma}=\left(\sigma_1,\sigma_2,\sigma_3\right)$ are the Pauli matrices:

\begin{equation}
\sigma_1=
\begin{pmatrix}
0 & 1 \\
1 & 0
\end{pmatrix},\;\;\;\;
\sigma_2=\begin{pmatrix}
0 & -i \\
i & 0
\end{pmatrix}\;\;\;\; \mathrm{and}\;\;\;\;
\sigma_3=\begin{pmatrix}
1 & 0 \\
0 & -1
\end{pmatrix}.
\end{equation}

After expanding the right-hand side of Eq. (\ref{com}), we finally arrive at a conservation relation of the kind

\begin{equation}
\frac{\partial\hat{\vec{A}}(\vec{r})}{\partial t}+\vec{\nabla}.\hat{\vec{T}}(\vec{r})=0,
\end{equation}

\noindent where $\hat{\vec{T}}(\vec{r})$ represents a tensorial operator whose elements read

\begin{equation}\label{tens}
\hat{T}_{ij}=\frac{c}{2}\left\{\hat{A}_i,\{\alpha_j,\delta(\vec{R}-\vec{r})\}\right\}.
\end{equation}

The development relies on the sum of operators of the kind $\hat{G}\delta \hat{D}$ such that

\begin{equation}
\langle\psi|\hat{G}\delta \hat{D}|\phi\rangle=\left(\hat{G}^+\psi^*(\vec{R})\right)\left(\hat{D}\phi(\vec{R})\right).
\end{equation}

The commutator of Eq. (\ref{com}) becomes an anti-commutator since $\hat{p}_j^*=-\hat{p}_j$. The elements $\hat{P}_{ij}$ of the pressure tensor are obtained by replacing the operator $\hat{\vec{A}}$ by the operator $\hat{\vec{p}}=\frac{\hbar}{i}\vec{\nabla}$ in Eq. (\ref{tens}); we get

\begin{equation}\label{pij}
\hat{P}_{ij}=\frac{c}{2}\left\{\frac{1}{i}\nabla_i,\{\alpha_j,\delta(\vec{R}-\vec{r})\}\right\}.
\end{equation}

Finally, the elements $\mathcal{P}_{ij}$ of the pressure tensor read

\begin{equation}
\mathcal{P}_{ij}=\sum_s\langle\psi_s|\hat{P}_{ij}|\psi_s\rangle,
\end{equation}

\noindent where $\psi_s$ are the quadri-vectors solutions of Dirac's equation. In the framework of the atom immersed in a spherical cavity, the pressure is given by the only element $\mathcal{P}_{rr}$ of the tensor, evaluated at the radius $R$ of the cavity \cite{MORE85}. After expanding the right-hand side of Eq. (\ref{pij}), we find

\begin{eqnarray}
\mathcal{P}_{rr}=\frac{c}{2i}\sum_s\left[-\left(\alpha_r\frac{\partial}{\partial r}\psi_s^*\right)\psi_s-\left(\frac{\partial\psi_s^*}{\partial r}\right)\alpha_r\psi_s+\left(\alpha_r\psi_s^*\right)\left(\frac{\partial\psi_s}{\partial r}\right)+\psi_s^*\left(\alpha_r\frac{\partial}{\partial r}\psi_s\right)\right]_R.
\end{eqnarray}

Using the equality

\begin{equation}
\sum_m\Omega_{j\ell m}^*(\theta,\phi)\Omega_{j\ell'm}(\theta,\phi)=\frac{2\ell+1}{4\pi}\delta_{\ell\ell'}
\end{equation}

\noindent and, following Eq. (\ref{psis}):

\begin{equation}
\alpha_r\psi_{j\ell m}=\left(
\begin{array}{l}
-\frac{1}{r}f(r)\Omega_{j\ell'm}(\theta,\phi)\\
\frac{i}{r}g(r)\Omega_{j\ell m}(\theta,\phi)
\end{array}
\right),
\end{equation}

\noindent the quantity $\mathcal{P}_{rr}$ can be put in a simple form, depending only on the radial components $f$ and $g$ and their derivatives, evaluated at the radius $R$ of the cavity:

\begin{equation}
\mathcal{P}_{rr}=-c\sum_j\frac{2j+1}{4\pi R^2}\left(f\frac{dg}{dr}-g\frac{df}{dr}\right)_R.
\end{equation}

A similar expression holds for the free states (with an integral over the free-electron energies). We show in Appendix A how the formula tends, in the non-relativistic limit, to the form provided by More \cite{MORE85}. Finally, after adding the contribution from exchange-correlation, $\nu$ and $V_{\infty}$ we get, for the total electron pressure

\begin{eqnarray}\label{presrel}
P_e&=&-c\sum_bn_b\frac{2j+1}{4\pi R^2}\left(f\frac{dg}{dr}-g\frac{df}{dr}\right)_R-c\sum_f\int_0^{\infty}n(\epsilon)\frac{2j+1}{4\pi R^2}\left(f\frac{dg}{dr}-g\frac{df}{dr}\right)_Rd\epsilon\nonumber\\
& &-\rho(R)f_{\mathrm{xc}}\left[\rho(R),T\right]-\nu\rho(R)-V_{\infty}\rho(R),
\end{eqnarray}

\noindent where $n_b$ and $n(\epsilon)$ are provided in Eq. (\ref{fdp}). The contributions from exchange and the quantities $V_{\infty}$ and $\nu$ equal the exchange-correlation pressure in the jellium of density $\bar{\rho}$:

\begin{equation}
P_{\mathrm{xc}}=-\bar{\rho}\left(f_{\mathrm{xc}}\left[\bar{\rho},T\right]-\mu_{\mathrm{xc}}\left[\bar{\rho},T\right]\right).
\end{equation}

\section{Stability of shock front in carbon, aluminum, silicon and niobium}\label{sec3}

The Rankine-Hugoniot relation for a planar shock reads

\begin{equation}\label{hug}
E-E_0=\frac{1}{2}\left(P+P_0\right)\left(\frac{1}{\rho_0}-\frac{1}{\rho}\right),
\end{equation}

\noindent where $\rho$, $P$, and $E$ represent respectively the density, pressure and internal energy of the shocked material while $\rho_0$, $P_0$ and $E_0$ are respectively the density, pressure and internal energy of the pole. The shock and material velocities $D$ and $u$ can be expressed as

\begin{equation}
D=\sqrt{\frac{\rho\left(P-P_0\right)}{\rho_0\left(\rho-\rho_0\right)}}
\end{equation}

\noindent and

\begin{equation}
u=\frac{P-P_0}{\rho_0D}.
\end{equation}

\subsection{Case of carbon}

\begin{figure}[H]
\begin{center}
\includegraphics[width=8.5cm]{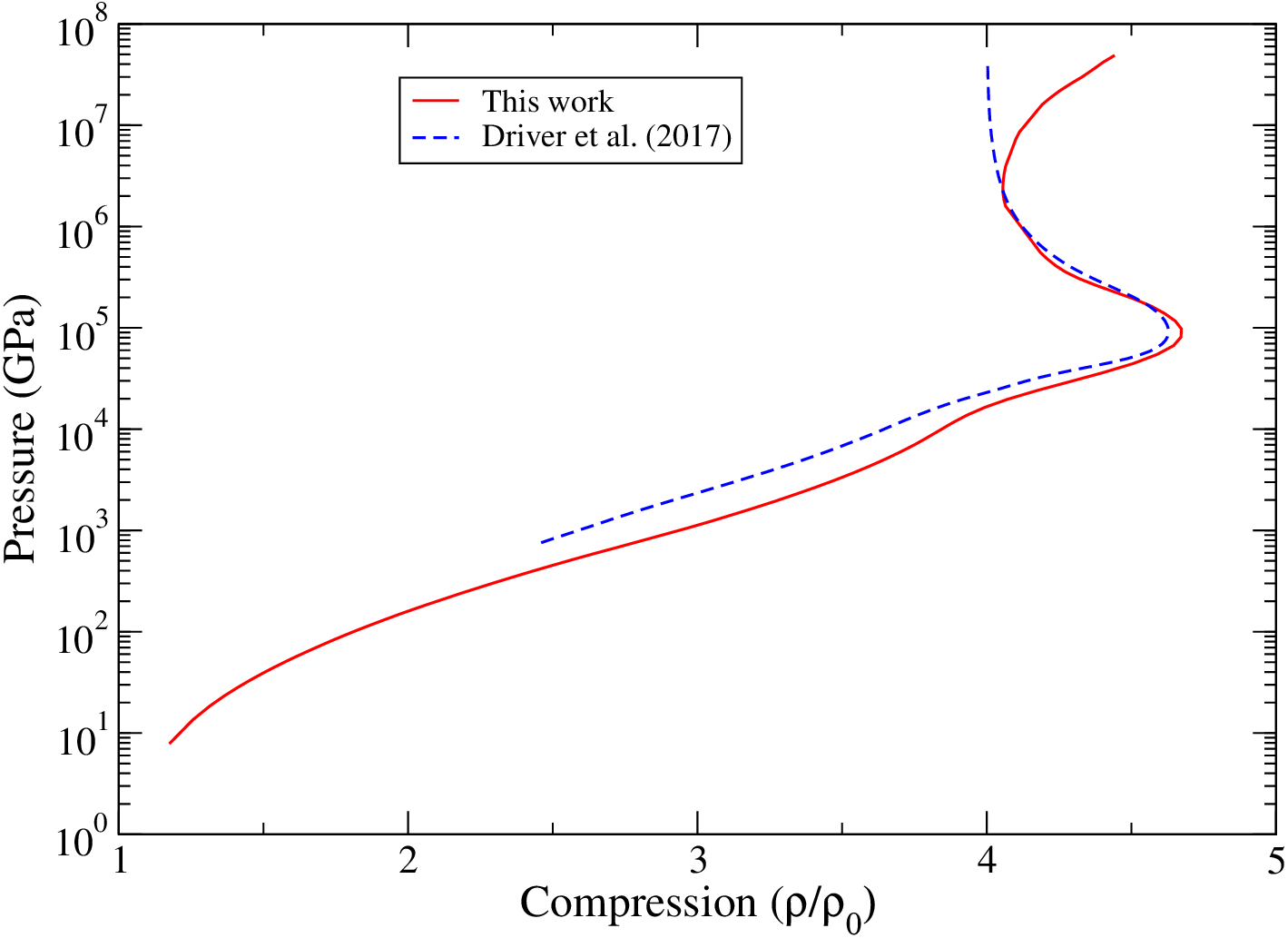}
\end{center}
\caption{Pressure along the principal Hugoniot of carbon. Comparison between our results (relativistic quantum average-atom model) and the Hugoniot obtained from Density Functional Theory-Molecular Dynamics (for $T<10^6$ K) simulations and Path Integral Monte Carlo (for $T>10^6$ K) by Driver \emph{et al.} \cite{DRIVER17}.}
\label{hug_C_withPIMC}
\end{figure}

\begin{figure}[H]
\begin{center}
\vspace{1cm}
\includegraphics[width=8.5cm]{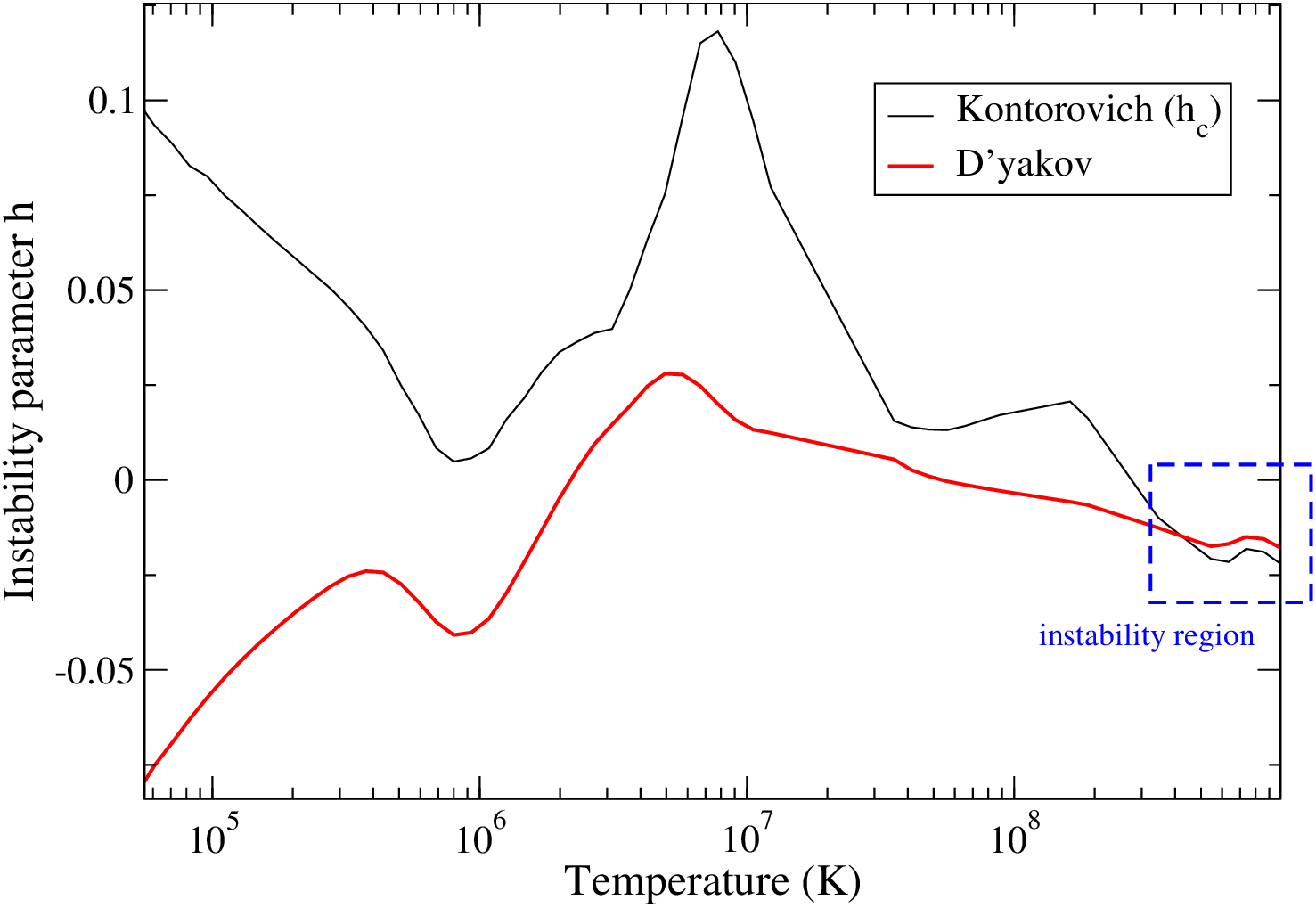}
\end{center}
\caption{Value of D'yakov's instability parameter $h$ \cite{DYAKOV54} along the principal Hugoniot of carbon, compared to the critical value $h_c$ as a function of temperature $T$.}
\label{h_C}
\end{figure}

\begin{figure}[H]
\begin{center}
\vspace{1cm}
\includegraphics[width=8.5cm]{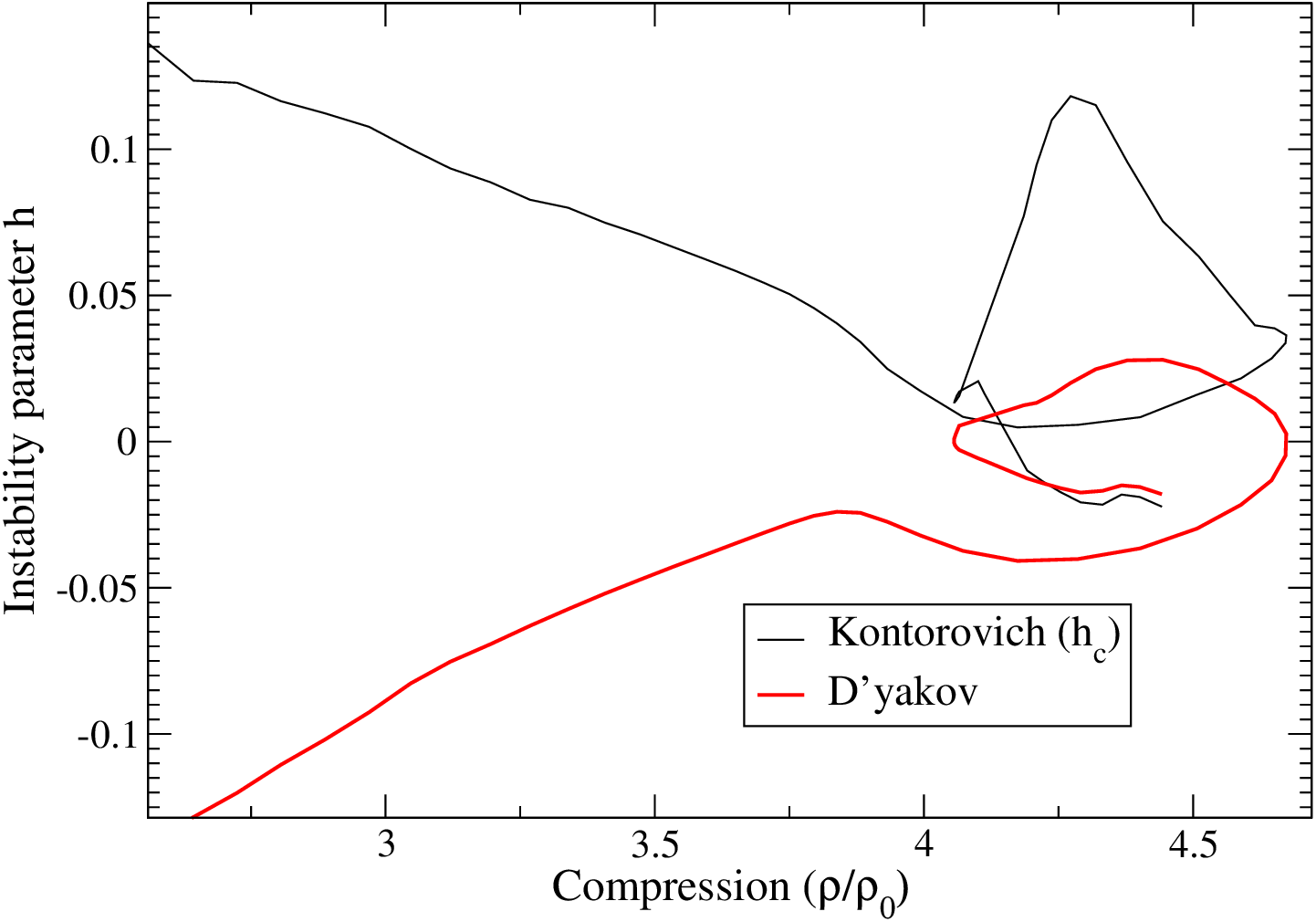}
\end{center}
\caption{Value of D'yakov's instability parameter $h$ \cite{DYAKOV54} along the principal Hugoniot of carbon, compared to the critical value $h_c$ as a function of compression $\rho/\rho_0$.}
\label{h_C_dens}
\end{figure}

\begin{figure}[H]
\begin{center}
\vspace{1cm}
\includegraphics[width=8.5cm]{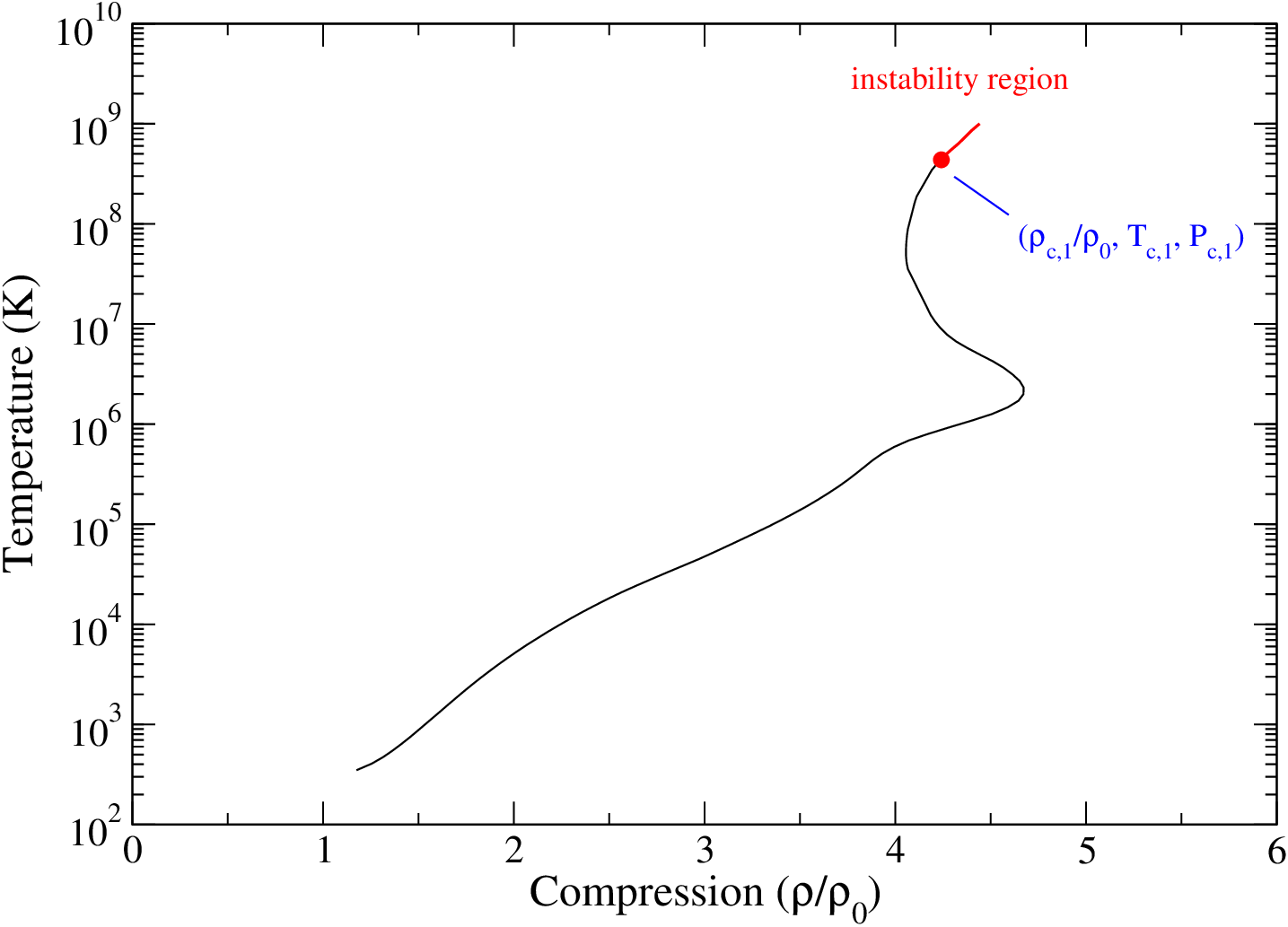}
\end{center}
\caption{Temperature and compression along the principal Hugoniot of carbon. The instability region ($h>h_c$) is indicated in red.}
\label{hug_rho_T_C}
\end{figure}

The initial density of carbon in the normal temperature and pressure conditions is chosen to be $\rho_0$=2.267 g/cm$^3$. Figure \ref{hug_C_withPIMC} shows the principal Hugoniot curve in the $\left(\rho/\rho_0, P\right)$ representation. The figure also displays, for comparison, the Hugoniot obtained from Density Functional Theory-Molecular Dynamics (DFT-MD) (for $T<$10$^6$ K) and Path Integral Monte Carlo (for $T>$10$^6$ K) by Driver \emph{et al.} \cite{DRIVER17}. The two shock adiabats are very close, also there is a small difference in the maximum compression, which is larger with our approach (by about 1 \%) in the region of ionization of the $L$ ($n$=1) shell. At high temperature, the DFT-MD / PIMC Hugoniot does not depart from the asymptote $\rho/\rho_0$=4 for tending to the asymptote $\rho/\rho_0$=7 because the simulations published in Ref. \cite{DRIVER17} are non-relativistic. Figure \ref{h_C} displays the value of D'yakov's instability parameter \cite{DYAKOV54} (see Eq. (\ref{dyako})) along the principal Hugoniot of carbon, compared to the critical value $h_c$ as a function of temperature $T$. The same two latter quantities are plotted versus compression $\rho/\rho_0$ in Fig. \ref{h_C_dens}. The correspondence between temperature and density along the principal Hugoniot path is provided by Fig. \ref{hug_rho_T_C}.

We can see that the instability occurs for temperatures above $T_{c,1}$=4.68$\times$10$^8$ K, which corresponds to compression $\rho_{c,1}/\rho_0$=4.37 and pressure $P_{c,1}$=2.19$\times$10$^7$ GPa. The latter compression is slightly greater than the $\rho/\rho_0$=4 asymptotic limit of the Hugoniot curve in the non-relativistic approximation (see Appendix B). In the relativistic case (Dirac equation for the electrons), the asymptote corresponds to compression 7 (see Appendix B), and compressions much larger than $4$ are then possible at high temperatures. This is the reason why, in our previous work \cite{HEUZE09}, we have not detected any instability, since the average-atom model we used relied on Schr\"odinger equation. The instability occurs at very high temperatures when the principal Hugoniot curve starts to depart from the non-relativistic asymptote $\rho/\rho_0$=4 to reach the asymptote $\rho/\rho_0$=7. Typically, the coupling parameter 

\begin{equation}\label{gam}
\Gamma=\frac{Z^{*2}e^2}{r_{\mathrm{ws}}k_BT}
\end{equation}

\noindent is around 0.05 and one can consider that the relativistic effects start to play an important role for $k_BT\geq 0.01\times c^2\approx 5$ keV. In the instability region, the ions are almost fully stripped and the most important contribution to the EOS is the electronic part.

\vspace{7mm}

\subsection{Case of aluminum}

\begin{figure}[H]
\begin{center}
\vspace{1cm}
\includegraphics[width=8.5cm]{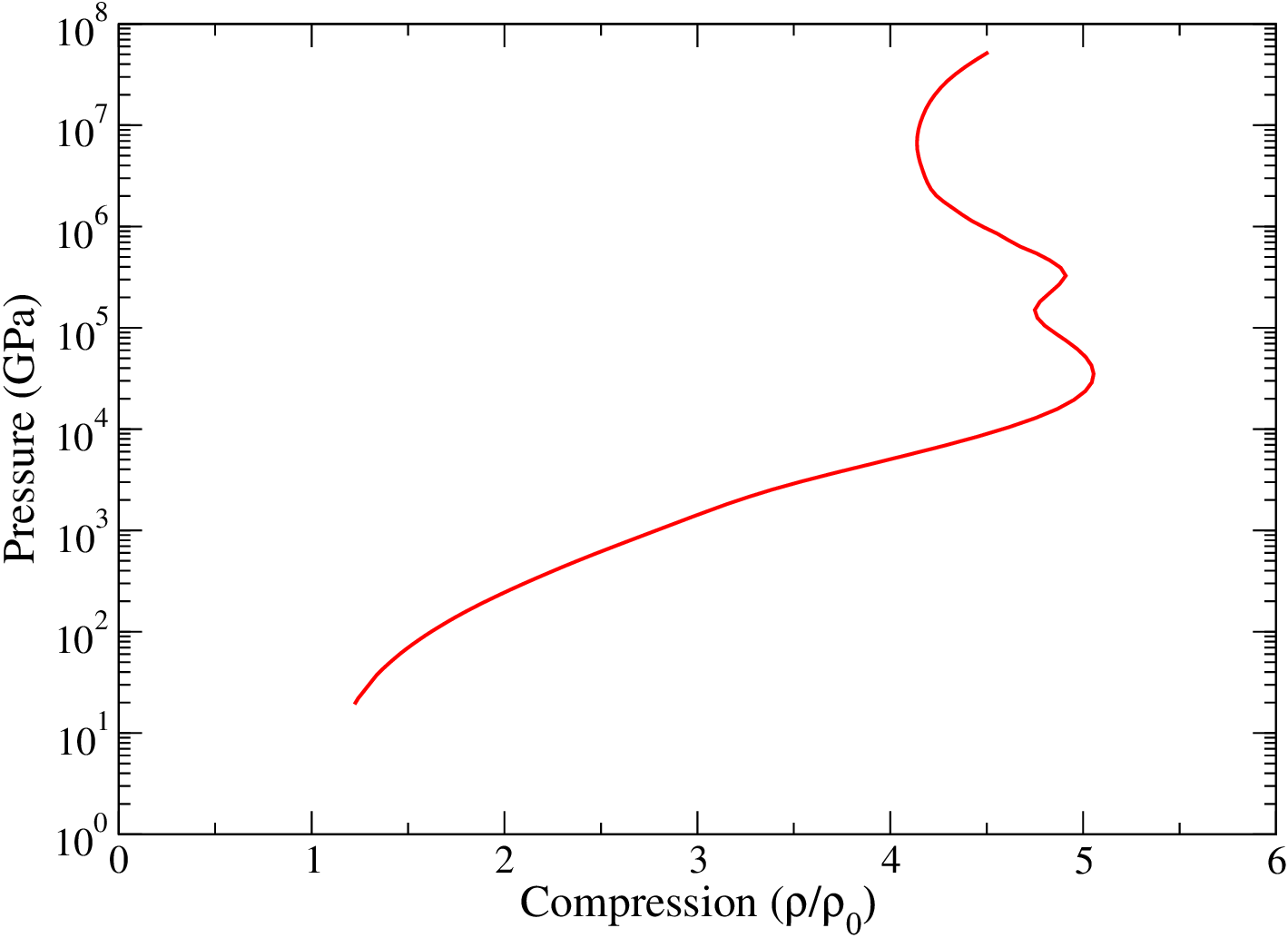}
\end{center}
\caption{Principal Hugoniot of aluminum.}
\label{hug_Al}
\end{figure}

\begin{figure}[H]
\begin{center}
\vspace{1cm}
\includegraphics[width=8.5cm]{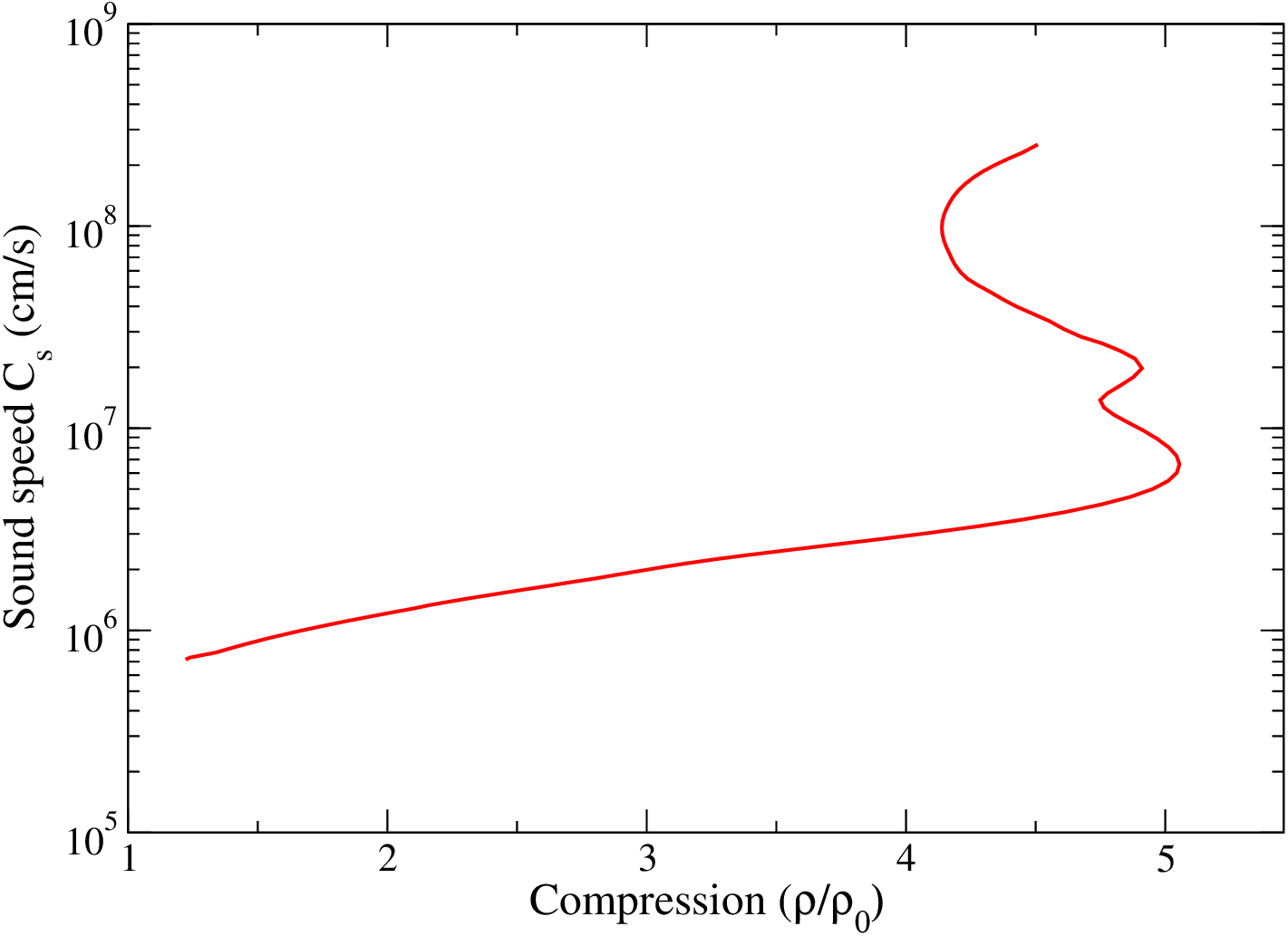}
\end{center}
\caption{Sound speed along the principal Hugoniot of aluminum.}
\label{cs_Al}
\end{figure}

The initial density of aluminum in the normal temperature and pressure conditions is chosen to be $\rho_0$= 2.7 g/cm$^3$. Figures \ref{hug_Al} and \ref{hug_rho_T_Al} show the principal Hugoniot curve respectively in the $\left(\rho/\rho_0, P\right)$ and $\left(\rho/\rho_0, T\right)$ representations. The ``shoulders'' at high compression are due to the successive ionization of the $K$ ($n$=1) and $L$ ($n$=2) shells \cite{PAIN07b}. As can be seen in Fig. \ref{cs_Al}, such bumps are also visible in the $\left(\rho/\rho_0, c_s\right)$ representation, $c_s$ being the sound speed.
 
\begin{figure}[H]
\begin{center}
\vspace{1cm}
\includegraphics[width=8.5cm]{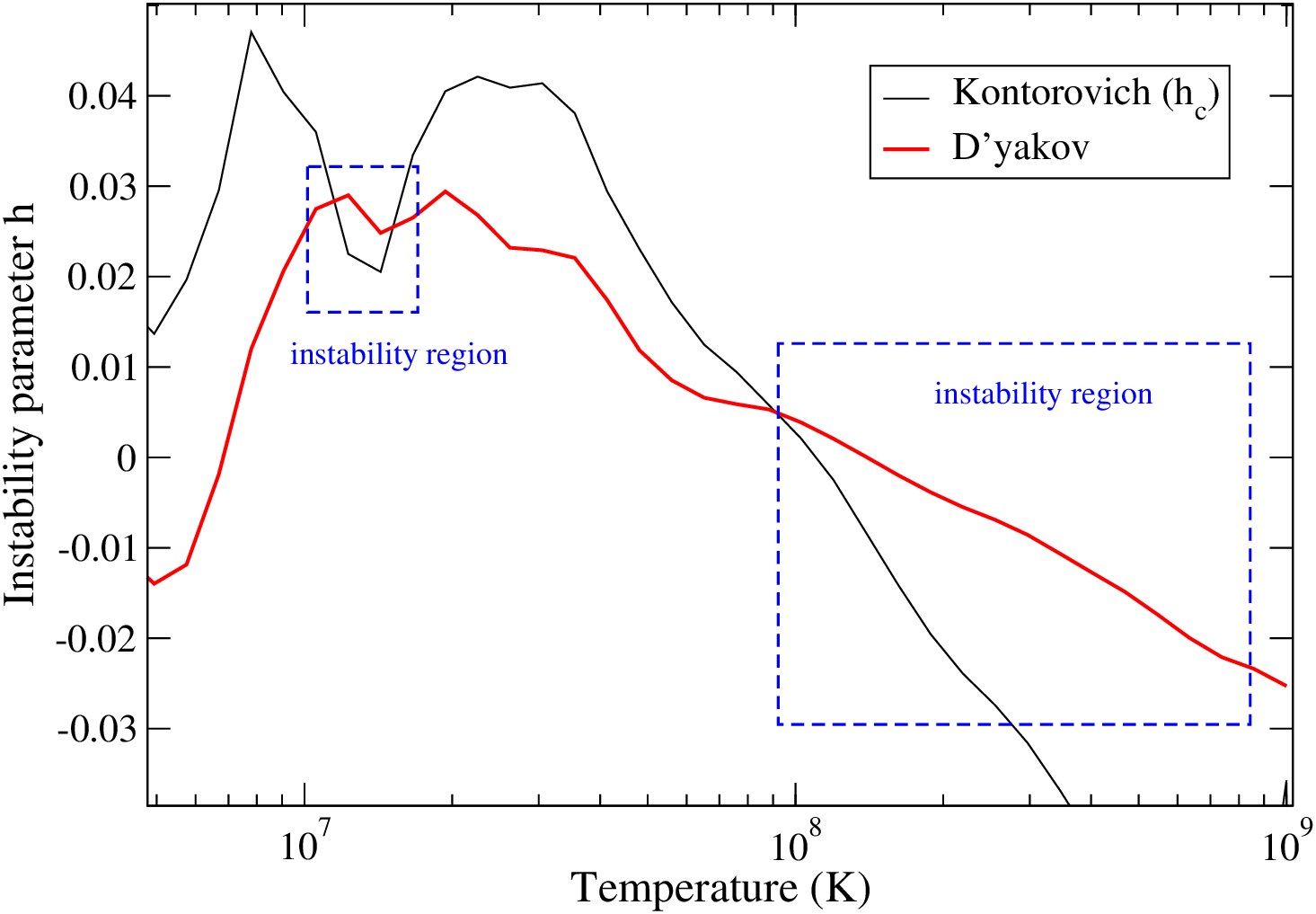}
\end{center}
\caption{Value of D'yakov's instability parameter $h$ \cite{DYAKOV54} along the principal Hugoniot of aluminum, compared to the critical value $h_c$ as a function of temperature $T$.}
\label{h_Al}
\end{figure}

\begin{figure}[H]
\begin{center}
\vspace{1cm}
\includegraphics[width=8.5cm]{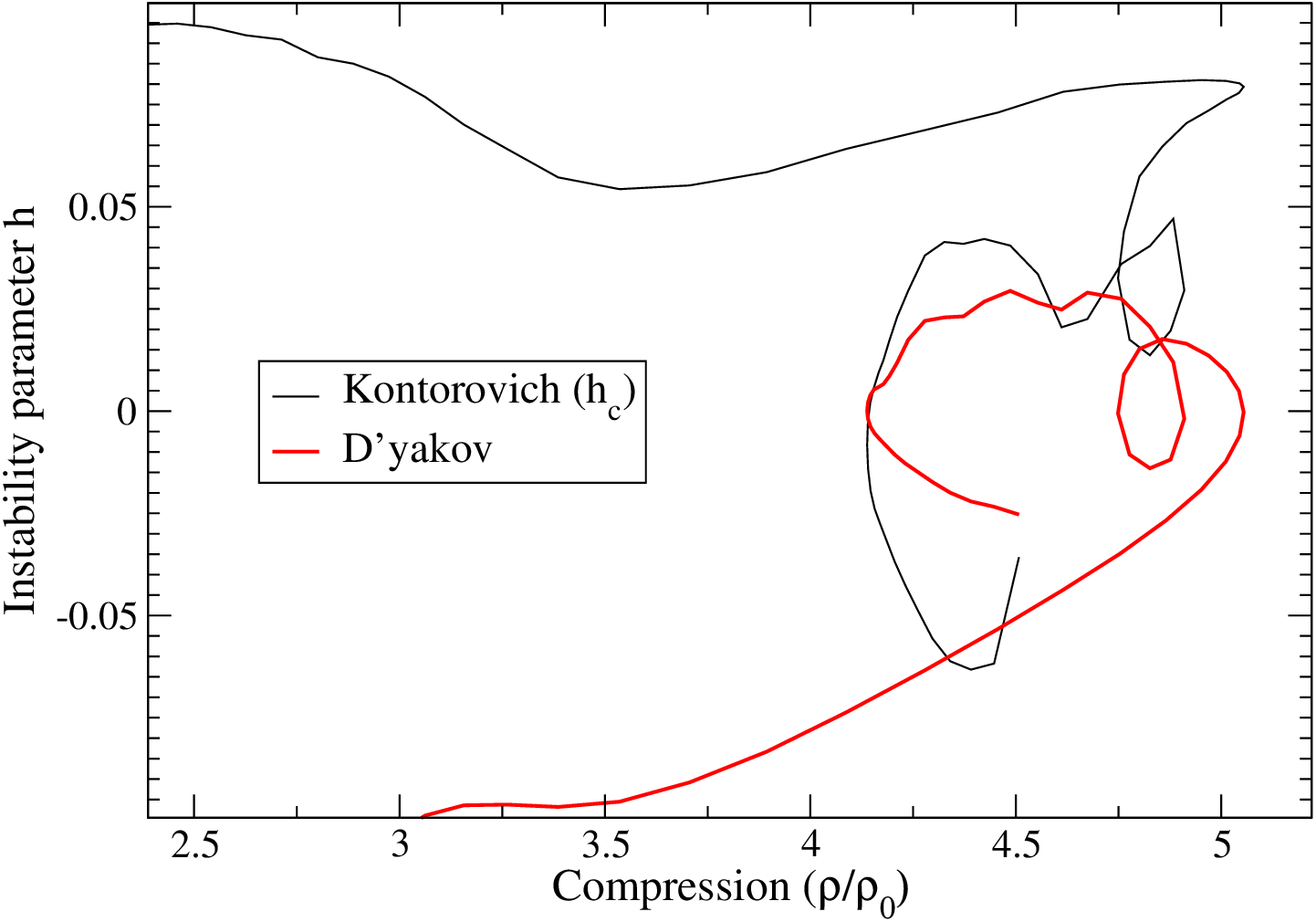}
\end{center}
\caption{Value of D'yakov's instability parameter $h$ \cite{DYAKOV54} along the principal Hugoniot of aluminum, compared to the critical value $h_c$ as a function of compression $\rho/\rho_0$.}
\label{h_Al_dens}
\end{figure}

\begin{figure}[H]
\begin{center}
\vspace{1cm}
\includegraphics[width=8.5cm]{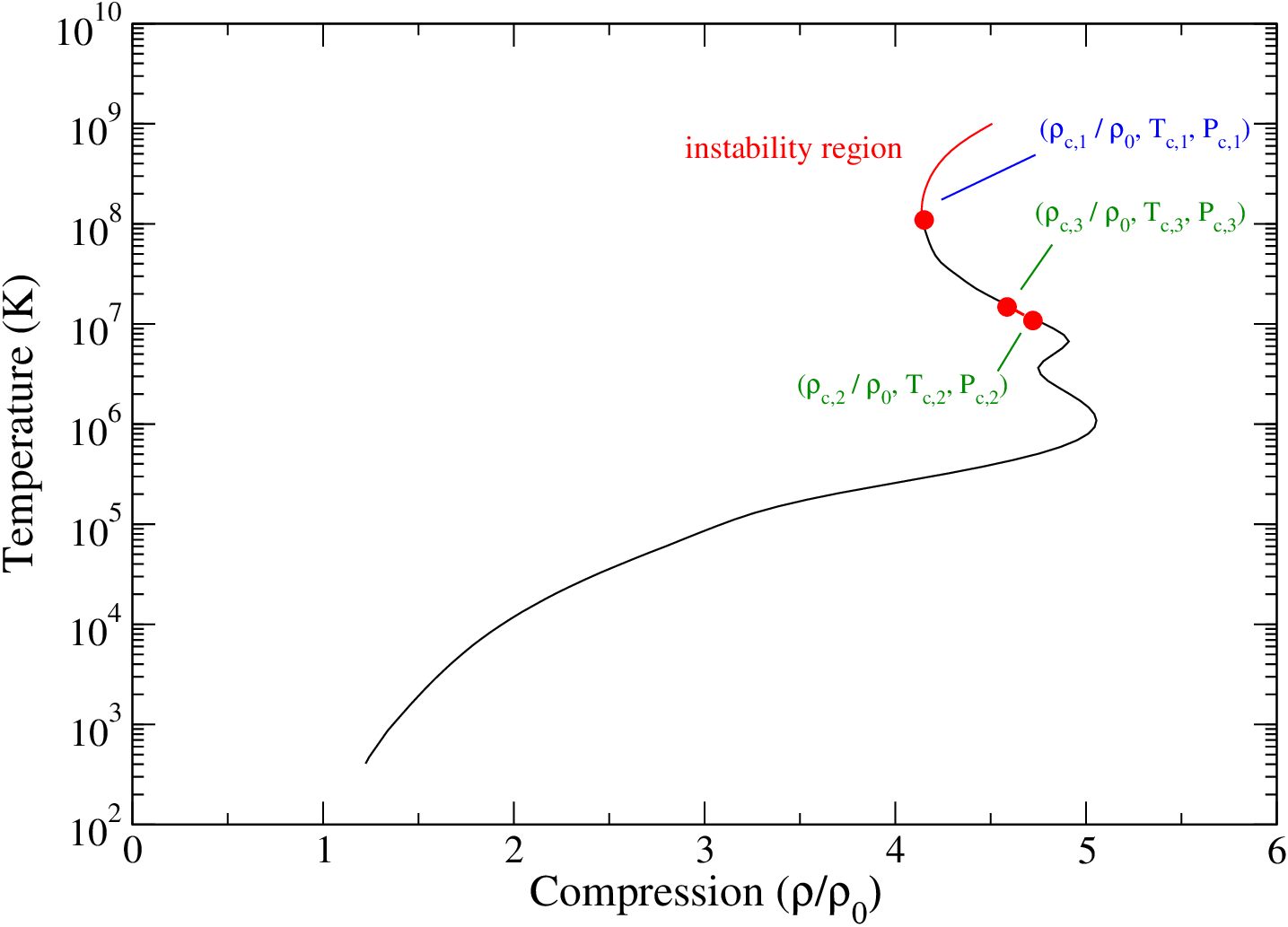}
\end{center}
\caption{Temperature and compression along the principal Hugoniot of aluminum. The instability region ($h>h_c$) is indicated in red.}
\label{hug_rho_T_Al}
\end{figure}

Figure \ref{h_Al} reveals that the instability ($h>h_c$) occurs for temperatures above $T_{c,1}$=1.03$\times$10$^8$ K, corresponding to $\left(\rho_{c,1}/\rho_0,P_{c,1}\right)$=(4.14, 4.96$\times$10$^6$ GPa). In such conditions, the relativistic effects start to play an important role, beyond the cusp point where the bifurcation towards compression 7 starts. This is confirmed by Fig. \ref{h_Al_dens}, where the D'yakov's parameter is plotted against compression. Typically, the coupling parameter in this instability region is around 0.3.

We also notice that the instability occurs as well for temperatures between $T_{c,2}$=1.22$\times$10$^7$ K and $T_{c,3}$=1.43$\times$10$^7$ K, corresponding to $\left(\rho_{c,2}/\rho_0,P_{c,2}\right)$=(4.67, 6.33$\times$10$^5$ GPa) and $\left(\rho_{c,3}/\rho_0,P_{c,3}\right)$=(4.61, 7.35$\times$10$^5$ GPa) respectively (the matching between temperature and density along the principal Hugoniot path is provided by Fig. \ref{hug_rho_T_Al}).

In the interesting study carried out by Das \emph{et al.} \cite{DAS11}, the authors find, for aluminum, two instability regions: the first one exists for a much lower temperature than we find (around $T\approx$ 150 eV, i.e., about 1.7$\times$10$^6$ K), after which D'yakov's parameter falls below the critical value $h_c$, and a second instability region starts around $T\approx$ 550 eV, i.e., 6.4$\times$10$^6$ K. Since this occurs in the conditions where ionization of $K$ and $L$ shells is important, we agree with the statement that D'yakov's instability is strongly connected to quantum electronic properties, but we find that the shock becomes unstable for much higher temperatures. This is probably due to the differences between our models (screened hydrogenic \emph{vs} quantum bound states and non-relativistic \emph{vs} relativistic). 

\subsection{Case of silicon}

\begin{figure}[H]
\begin{center}
\vspace{1cm}
\includegraphics[width=8.5cm]{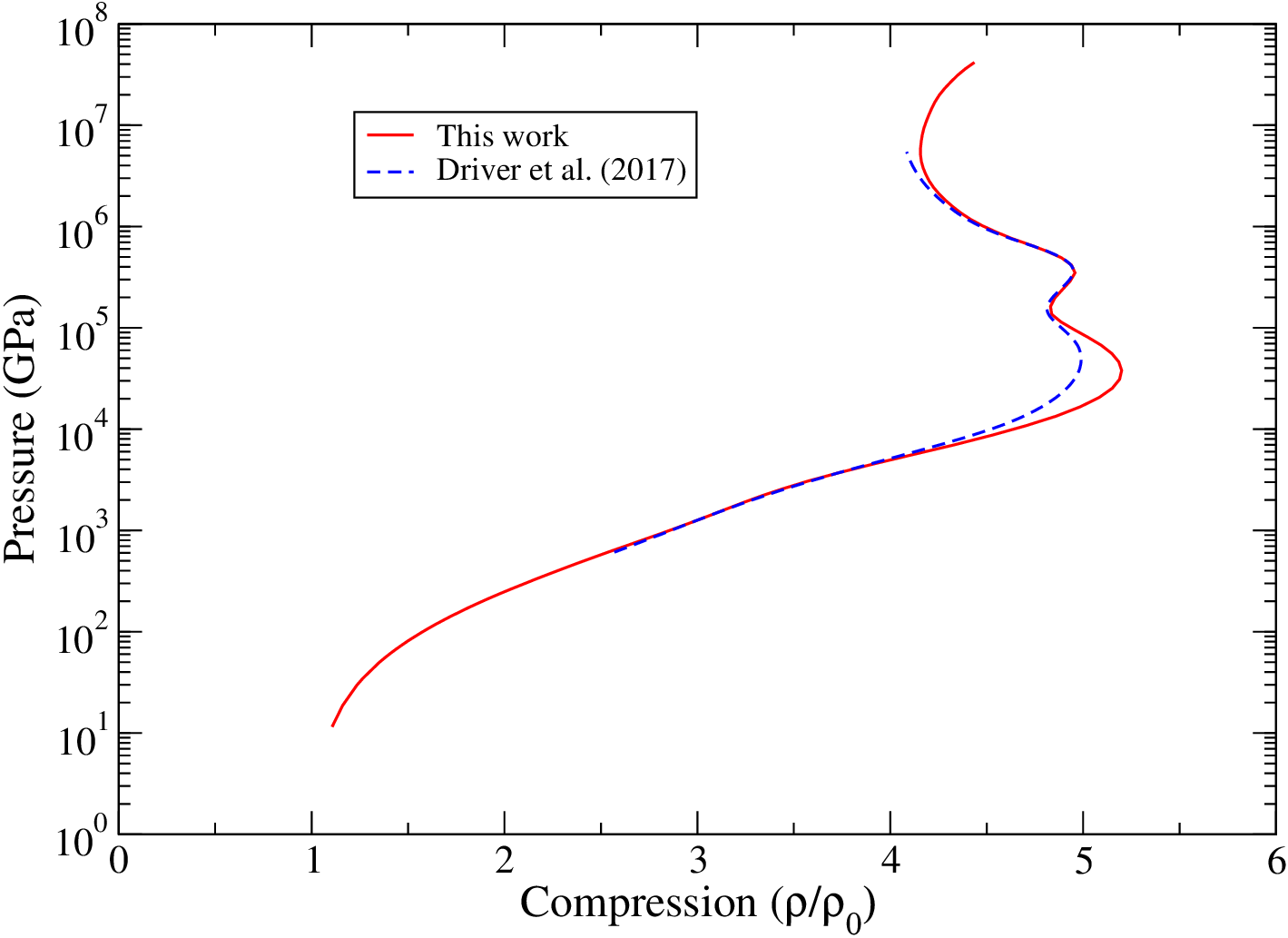}
\end{center}
\caption{Principal Hugoniot of silicon. Comparison between our results (realtivistic quantum average-atom model) and the Hugoniot obtained from Density Functional Theory-Molecular Dynamics (for $T<$10$^6$ K) simulations and Path Integral Monte Carlo (for $T>$10$^6$ K) by Driver \emph{et al.} \cite{DRIVER17}.}
\label{hug_Si_new_withPIMC_new}
\end{figure}

The initial density of silicon in the normal temperature and pressure conditions is chosen to be $\rho_0$= 2.33 g/cm$^3$. Figure \ref{hug_Si_new_withPIMC_new} displays our Hugoniot curve and the one obtained from PIMC / DFT-MD computations by Driver \emph{et al.}. As for carbon, our approach predicts a higher maximal compression (about 4 \%). 
 
\begin{figure}[H]
\begin{center}
\vspace{1cm}
\includegraphics[width=8.5cm]{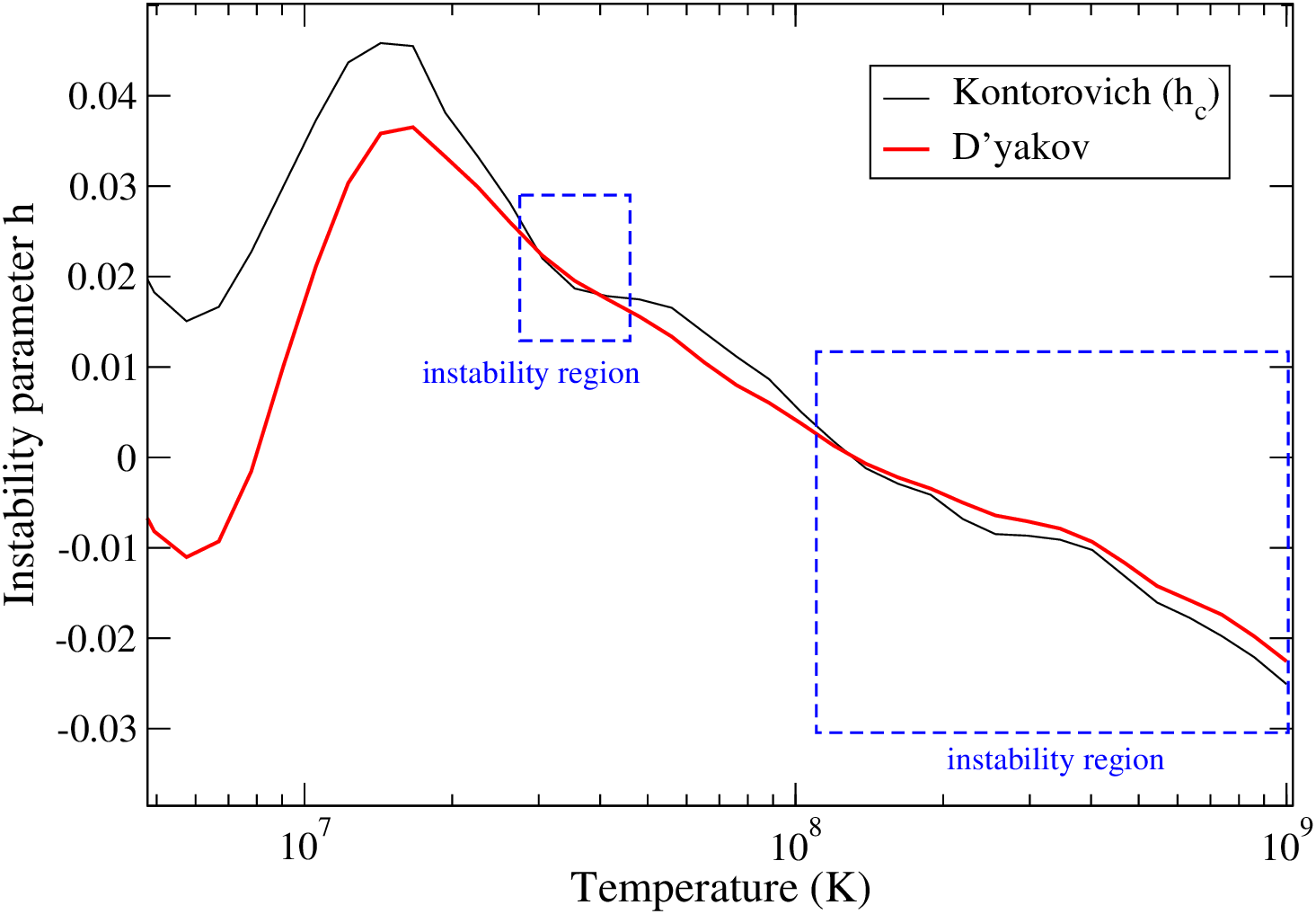}
\end{center}
\caption{Value of D'yakov's instability parameter $h$ \cite{DYAKOV54} along the principal Hugoniot of silicon, compared to the critical value $h_c$ as a function of temperature $T$.}
\label{h_Si}
\end{figure}

\begin{figure}[H]
\begin{center}
\vspace{1cm}
\includegraphics[width=8.5cm]{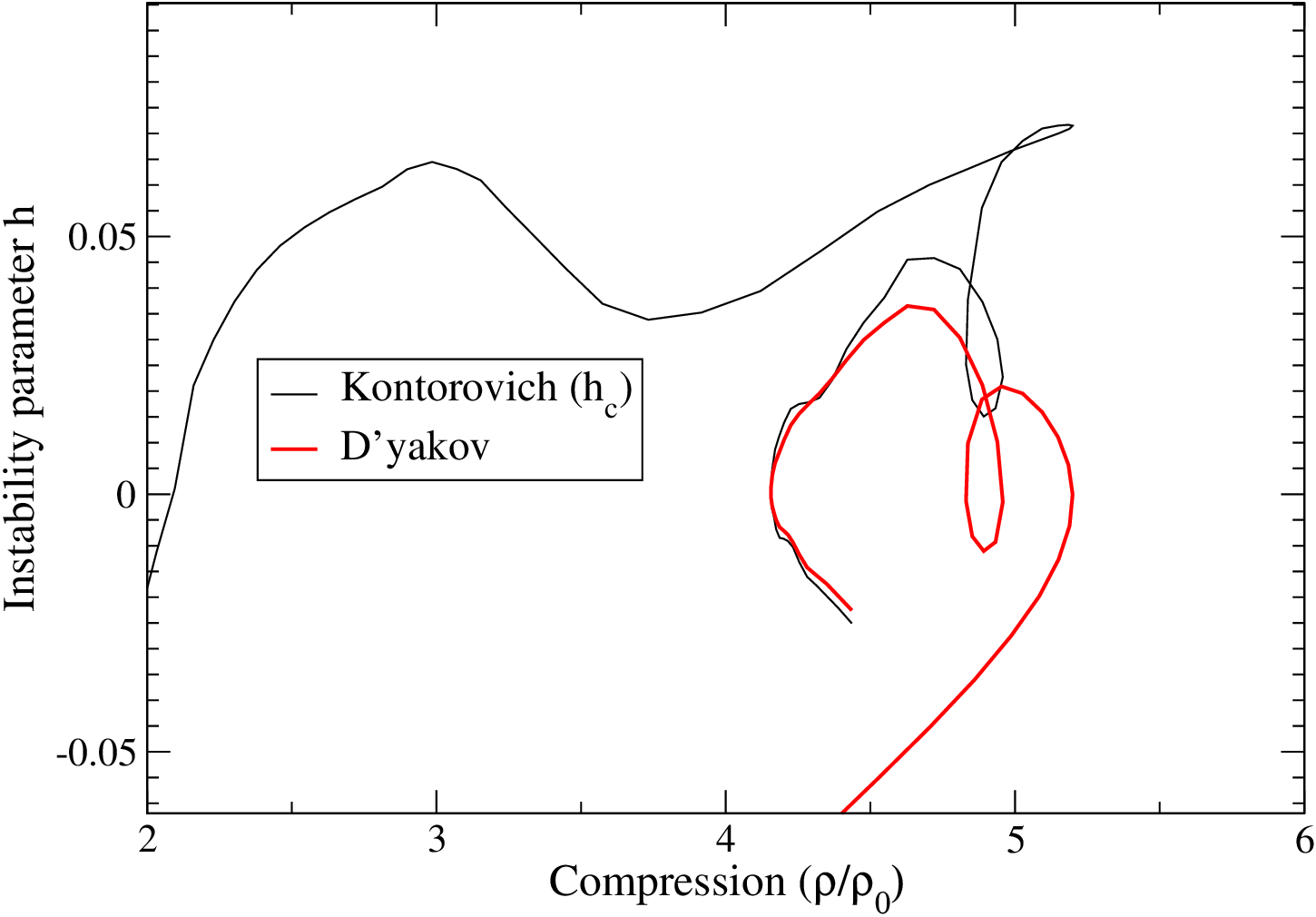}
\end{center}
\caption{Value of D'yakov's instability parameter $h$ \cite{DYAKOV54} along the principal Hugoniot of silicon, compared to the critical value $h_c$ as a function of compression $\rho/\rho_0$.}
\label{h_Si_dens}
\end{figure}

\begin{figure}[H]
\begin{center}
\vspace{1cm}
\includegraphics[width=8.5cm]{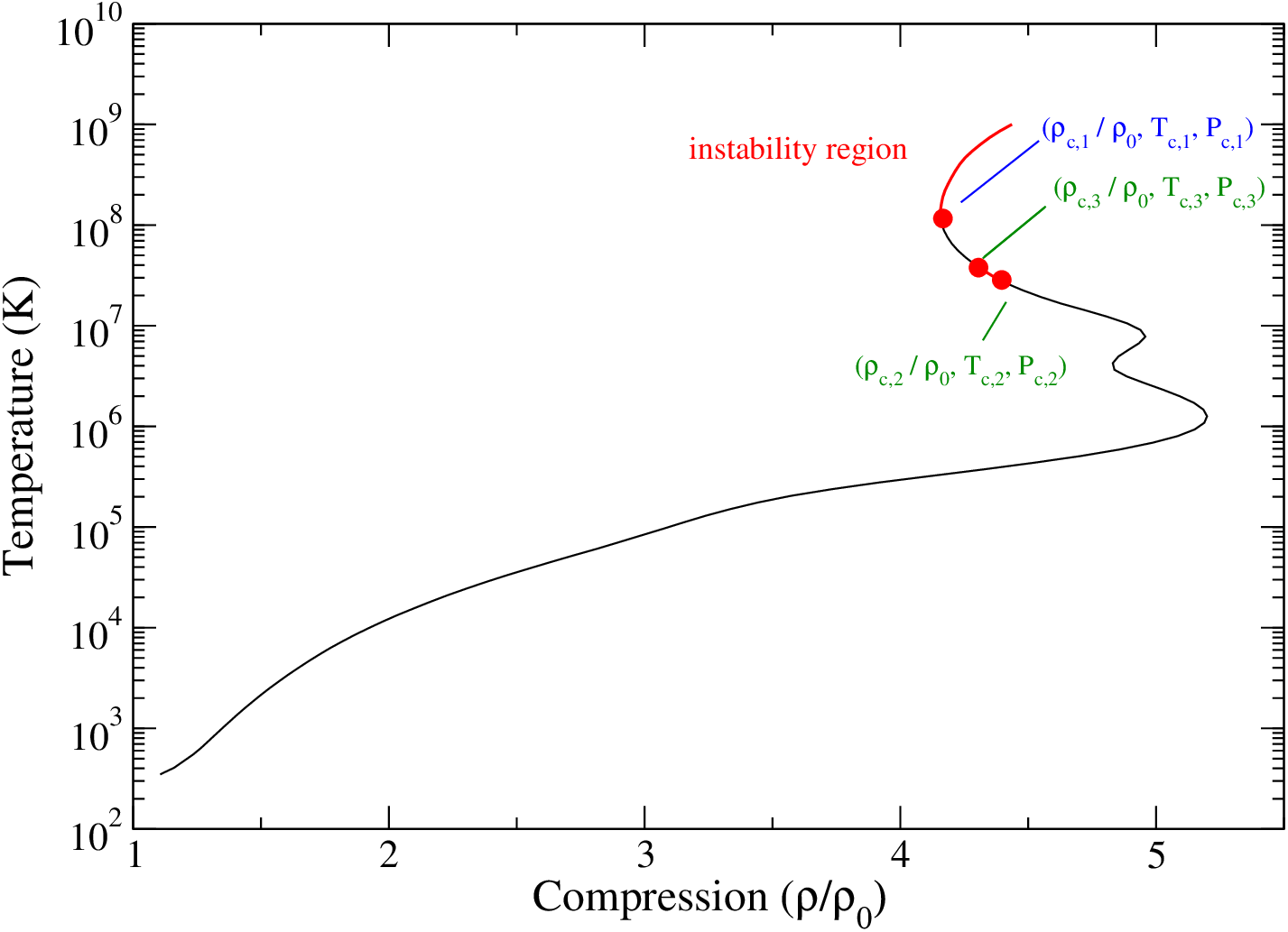}
\end{center}
\caption{Temperature and compression along the principal Hugoniot of silicon. The instability region ($h>h_c$) is indicated in red.}
\label{hug_rho_T_Si}
\end{figure}

Figure \ref{h_Si} indicates that the instability ($h>h_c$) occurs for temperatures above $T_{c,1}$=1.39$\times$10$^8$ K, corresponding to $\left(\rho_{c,1}/\rho_0,P_{c,1}\right)$=(4.16, 5.93$\times$10$^6$ GPa), after the departure from compression 4 to compression 7. The D'yakov's parameter is plotted against compression in Fig. \ref{h_Si_dens} and the temperature range in which the system becomes unstable can be viewed in Fig. \ref{hug_rho_T_Si}. This is not surprising since the atomic numbers of aluminium and silicon differ only by one. Typically, the coupling parameter in the instability region is around 0.3.

As for aluminum, we find a narrow region of instability between $T_{c,1}$=2.97$\times$10$^7$ K and $T_{c,2}$=4$\times$10$^7$ K, corresponding to $\left(\rho_{c,2}/\rho_0,P_{c,2}\right)$=(4.36, 1.36$\times$10$^6$ GPa) and $\left(\rho_{c,3}/\rho_0,P_{c,3}\right)$=(4.29, 1.81$\times$10$^6$ GPa) respectively (the matching between temperature and density along the principal Hugoniot path is provided by Fig. \ref{hug_rho_T_Si}).

\subsection{Case of niobium}

\begin{figure}[H]
\begin{center}
\vspace{1cm}
\includegraphics[width=8.5cm]{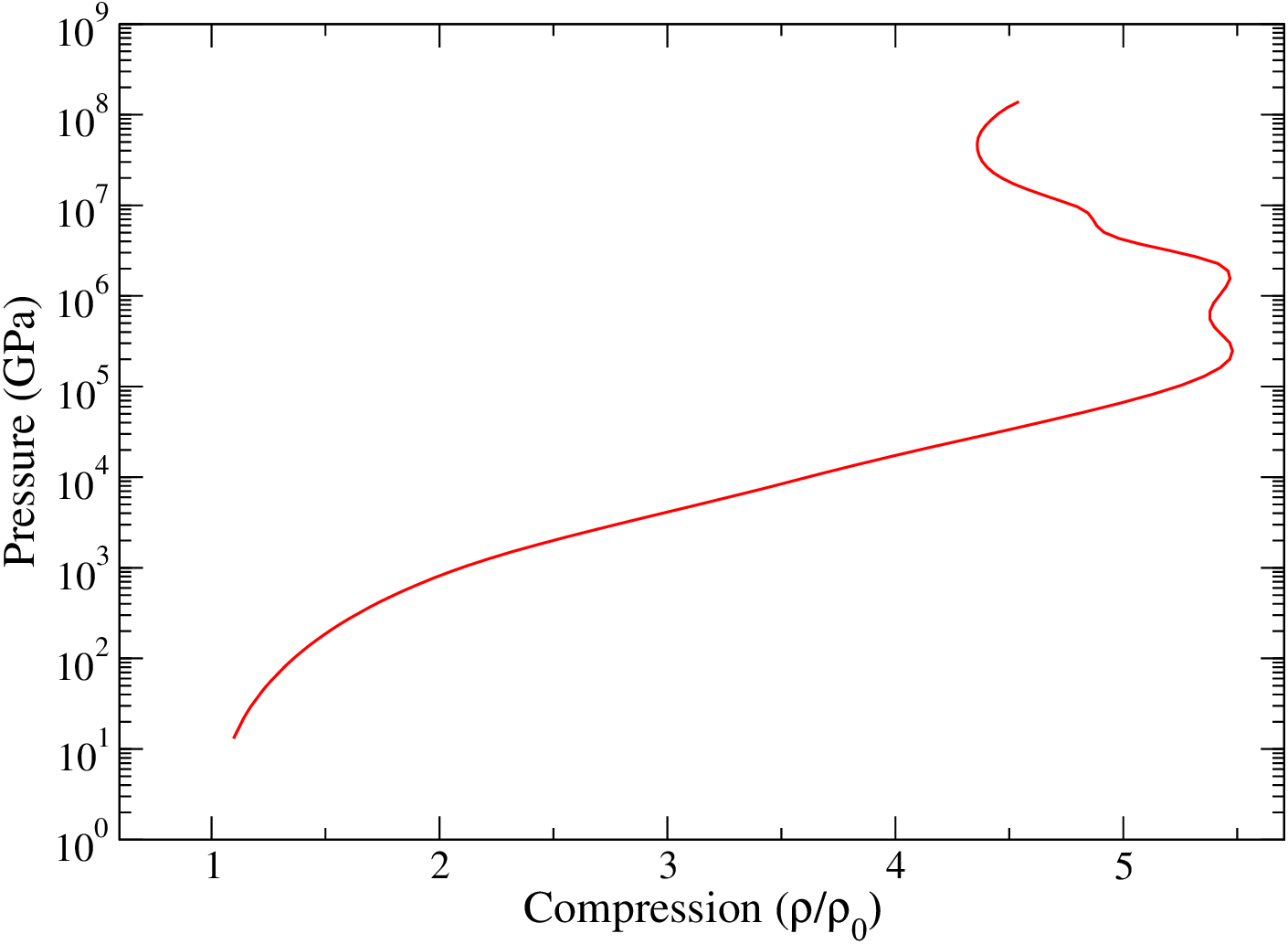}
\end{center}
\caption{Principal Hugoniot of niobium.}
\label{hug_Nb}
\end{figure}

\begin{figure}[H]
\begin{center}
\vspace{1cm}
\includegraphics[width=8.5cm]{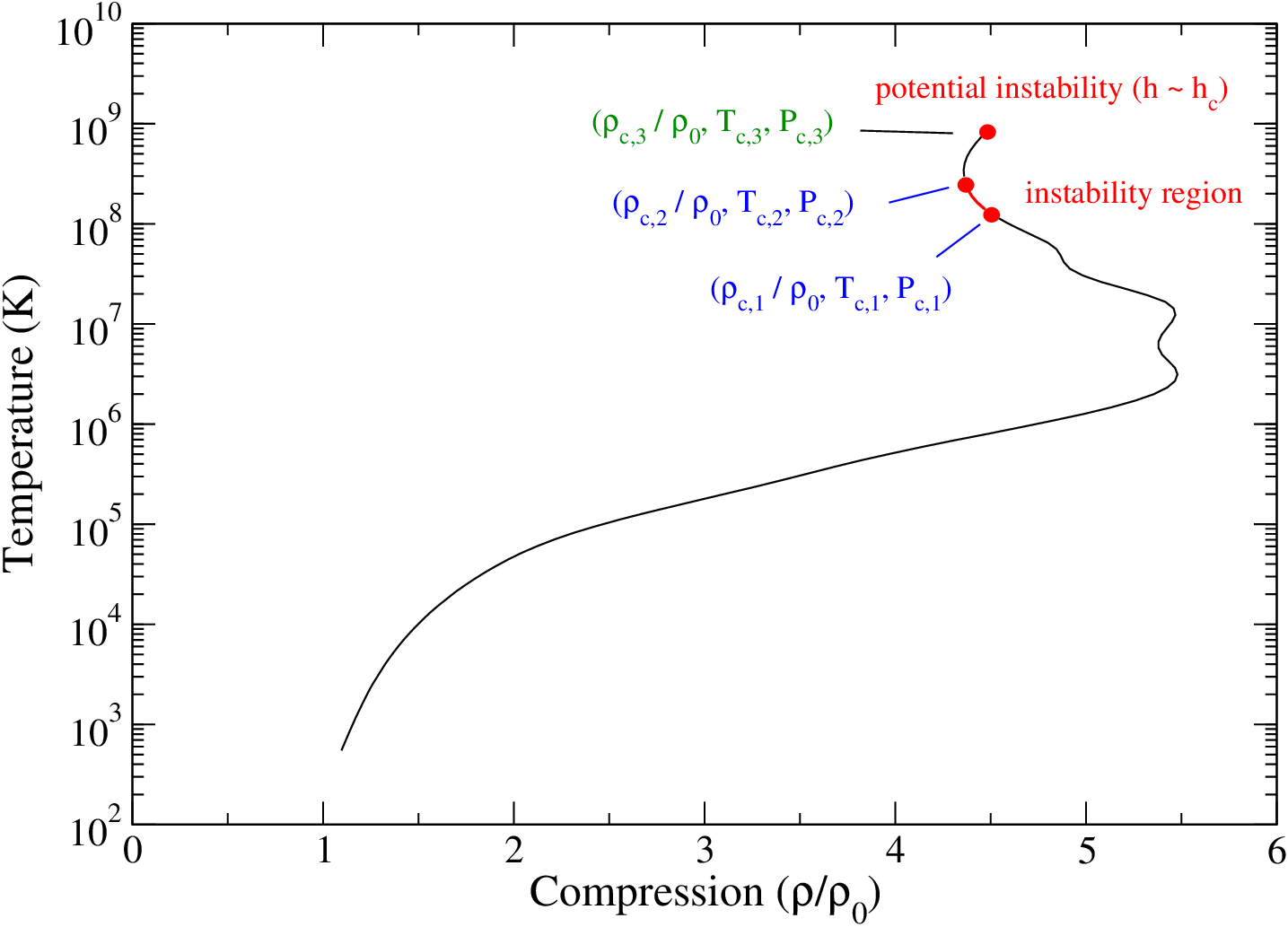}
\end{center}
\caption{Value of D'yakov's instability parameter $h$ \cite{DYAKOV54} along the principal Hugoniot of niobium, compared to the critical value $h_c$ as a function of compression $\rho/\rho_0$.}
\label{hug_rho_T_Nb}
\end{figure}

\begin{figure}[H]
\begin{center}
\vspace{1cm}
\includegraphics[width=8.5cm]{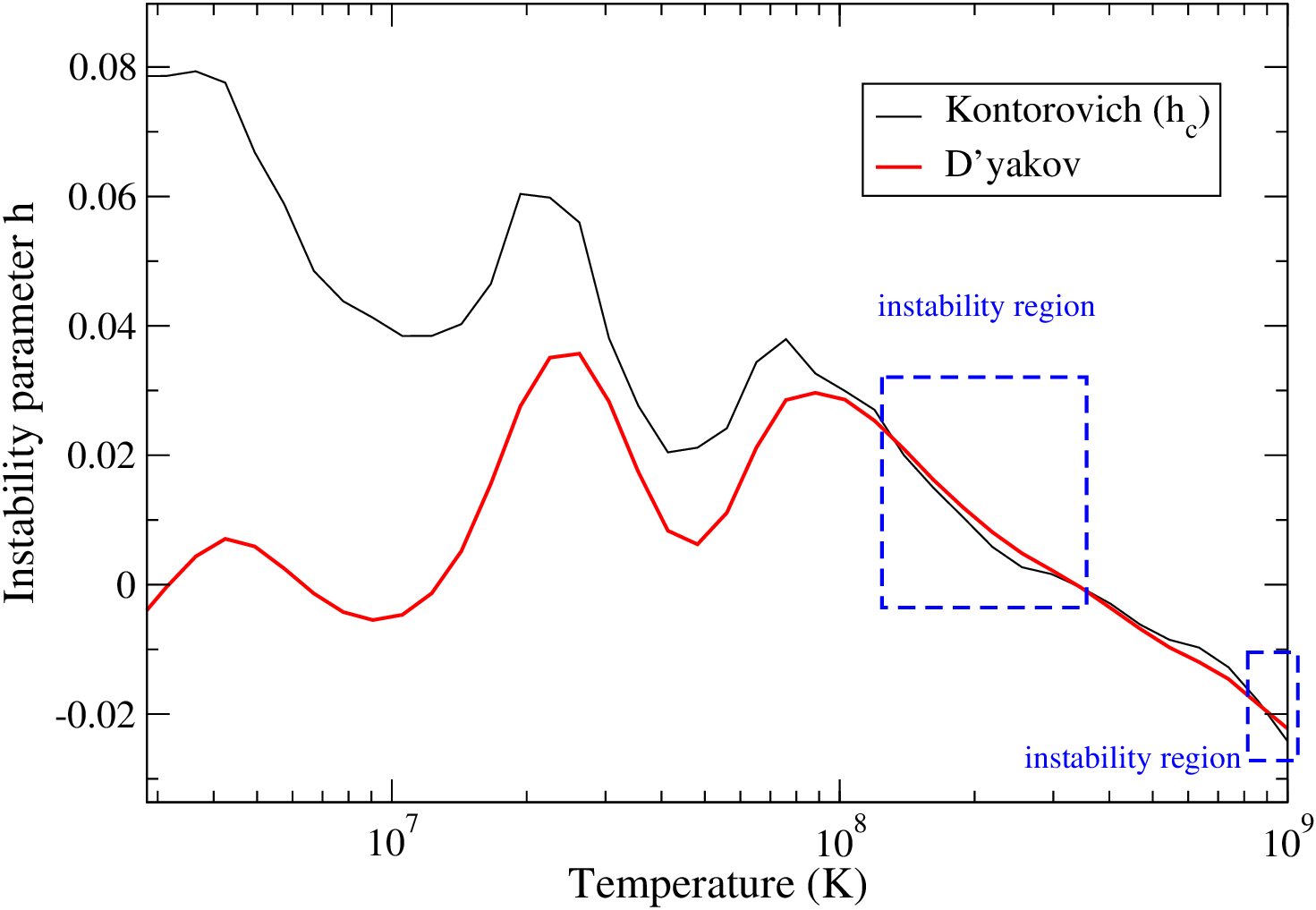}
\end{center}
\caption{Value of D'yakov's instability parameter $h$ \cite{DYAKOV54} along the principal Hugoniot of niobium, compared to the critical value $h_c$ as a function of temperature $T$.}
\label{h_Nb}
\end{figure}

\begin{figure}[H]
\begin{center}
\vspace{1cm}
\includegraphics[width=8.5cm]{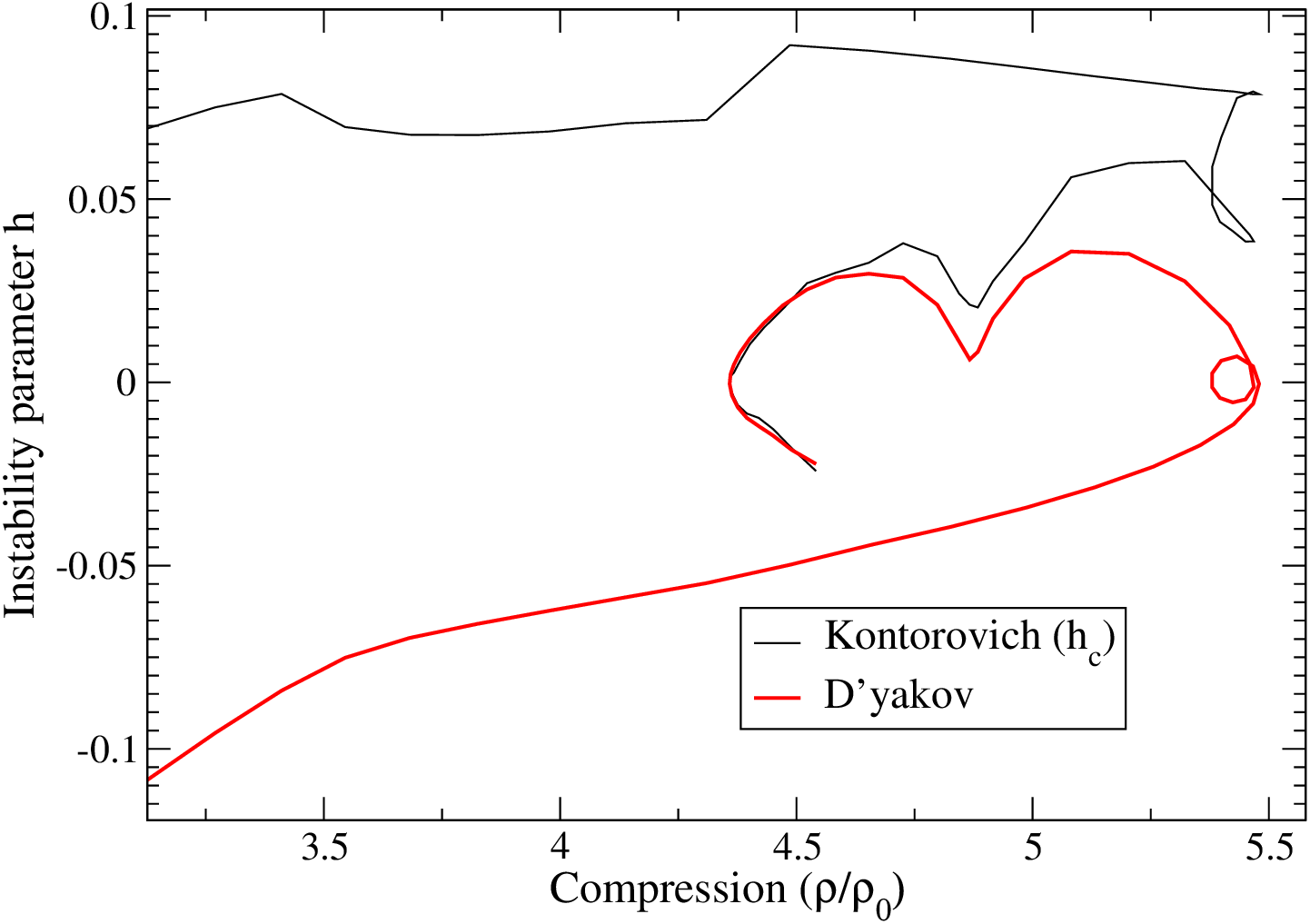}
\end{center}
\caption{Value of D'yakov's instability parameter $h$ \cite{DYAKOV54} along the principal Hugoniot of niobium, compared to the critical value $h_c$ as a function of compression $\rho/\rho_0$.}
\label{h_Nb_dens}
\end{figure}

The initial density of niobium in the normal temperature and pressure conditions is chosen to be $\rho_0$=8.57 g/cm$^3$. Figures \ref{hug_Nb} and \ref{hug_rho_T_Nb} display the principal Hugoniot curve respectively in the $\left(\rho/\rho_0, P\right)$ and $\left(\rho/\rho_0, T\right)$ representations. The D'yakov's instability parameter \cite{DYAKOV54} and the critical value $h_c$ along the principal Hugoniot of niobium are plotted as a function of temperature $T$ in Fig. \ref{h_Nb}, and versus compression $\rho/\rho_0$ in Fig. \ref{h_Nb_dens}. 

The instability occurs for temperatures between $T_{c,1}$=1.39$\times$10$^8$ K (compression $\rho_{c,1}/\rho_0$=4.47 and $P_{c,1}$=1.97$\times$10$^7$ GPa) and $T_{c,2}$=2.97$\times$10$^8$ K, for which the compression is $\rho_{c,2}/\rho_0$=4.36 and the pressure $P_{c,2}$=4.11$\times$10$^7$ GPa. As for carbon and silicon, the compression in that range is greater than the non-relativistic $\rho/\rho_0$=4 asymptotic limit of the Hugoniot curve. Here also, the instability occurs at very high temperatures just before the principal Hugoniot curve starts to depart from the $\rho/\rho_0$=4 asymptote to reach the limit $\rho/\rho_0$=7. Typically, the coupling parameter in the instability region is around 2-3. In that case the non-ideality correction to the ionic EOS (see Eqs. (\ref{iocp}) and (\ref{ig3})) are important.

At the end of the Hugoniot curve, i.e., at $T_{c,3}$=10$^9$ K, where compression is $\rho_{c,3}/\rho_0$=4.54 and pressure $P_{c,3}$=1.39$\times$10$^8$, we see also the beginning of another instability region.

\section{Radiative pressure and internal energy}\label{sec4}

Most of the earlier work on the subject \cite{BATES00,BATES12,BATES15} was focused on the low-temperature range. Mond \emph{et al.} \cite{MOND97} studied shock waves propagating in room-temperature argon, with post-shock temperatures not exceeding several eV. Konyukhov \emph{et al.} who approximated a realistic EOS for magnesium with a van der Waals model, found the D'yakov-Kontorovich unstable regions at plasma temperatures below 20 eV \cite{KONYUKHOV09}. At such low temperatures, one can safely neglect the energy density and pressure of the plasma radiation. The shock waves studied here are so strong that the post-shock plasma in thermodynamic equilibrium must be radiation-dominated. However, it is not clear to us whether the plasma radiation should be accounted for or not in the Hugoniot energy balance. For instance, Das \emph{et al.} \cite{DAS11} did not include it in their study of D'yakov-Kontorovich instability. This is justified because their temperatures are lower than 1.1 keV$\approx$ 1.28$\times$10$^7$ K, which corresponds to the end of the signature of the shell structure, before the relativistic effects start to play a role. The impact on the radiation field is important in a radiative-hydrodynamics simulation, and is different at each time step and spatial mesh. Usually, in ICF, its distribution is far from a Planckian. In the diffusion approximation, the one-dimensional radiative transfer equation reads

\begin{equation}
\frac{dI_{\nu}}{dx}=-\kappa_{\nu}I_{\nu}+j_{\nu},
\end{equation}

\noindent $I_{\nu}$ being the intensity of the radiation, $\kappa_{\nu}$ its opacity and $j_{\nu}$ its emissivity. The solution of such an equation, for an homogeneous medium and assuming that $\kappa_{\nu}$ and $j_{\nu}$ are related by Kirchoff's law

\begin{equation}
j_{\nu}=B_{\nu}\kappa_{\nu},
\end{equation}

\noindent where $B_{\nu}$ is Planck's distribution function, is given by

\begin{equation}
I_{\nu}=B_{\nu}\times\left[1-e^{-\rho L\kappa_{\nu}}\right],
\end{equation}

\noindent where $\rho$ is the density of the material and $L$ its thickness. If $\rho L\kappa_{\nu}\gg1$, the plasma is optically thick and one has $I_{\nu}=B_{\nu}$ and the radiation is the one of a blackbody.

\vspace{5mm}

In Sec. \ref{sec3}, we have not accounted for the energy and pressure of the equilibrium radiation when calculating the Hugoniot shock adiabats. This means that we have assumed implicitly that the shocked plasma is optically thin to its own radiation. Unlike the optically thick plasma, an optically thin plasma cannot be in a steady state; it is radiatively cooled, losing its thermal energy with time. To justify the steady-state Hugoniot calculations that ignore the cooling, one has to consider a sufficiently small depth of the shocked plasma layer and that the radiation energy loss can be neglected while the plasma flows through it. On the other hand, to apply the D'yakov-Kontorovich theory, this depth should be much larger than both the shock width and the transverse perturbation wavelength.

\vspace{5mm}

In the case where the plasma is optically thick, it may be relevant to include radiation in the EOS of the material when computing the Hugoniot \cite{ZHANG17}, through a blackbody radiation EOS model, as used for instance by Das and Menon \cite{DAS09}. This should give insight, at least qualitatively, into the impact of radiation on the instability criterion. The electrons in the plasma produce radiation heating via Bremsstrahlung and also reach equilibrium with the plasma via inverse Bremsstrahlung and Compton scattering process. As pointed out by Das and Menon \cite{DAS09}, the time scale for the equilibration of radiation with matter is generally of the order of $10^{-13}$ to $10^{-15}$ s, which is negligible in comparison to the time scale of shock propagation $10^{-9}$ to $10^{-6}$ s and hence justifies that the matter exists in equilibrium with radiation. Under equilibrium condition, the energy density of radiation depends on the temperature of the material. When an intense shock is launched, the temperature becomes so high that the energy density and pressure of radiation become comparable to the internal energy and pressure of electrons, thereby affecting both the EOS and Hugoniot. The derivation of the Hugoniot relations assumes that the post-shock matter is in full thermodynamic equilibrium, that is, it must be optically thick to its own radiation. The local thermodynamic equilibrium of the plasma with the radiation implies a Planckian spectrum corresponding to the post-shock temperature.

\clearpage

\begin{figure}[H]
\begin{center}
\vspace{1cm}
\includegraphics[width=8.5cm]{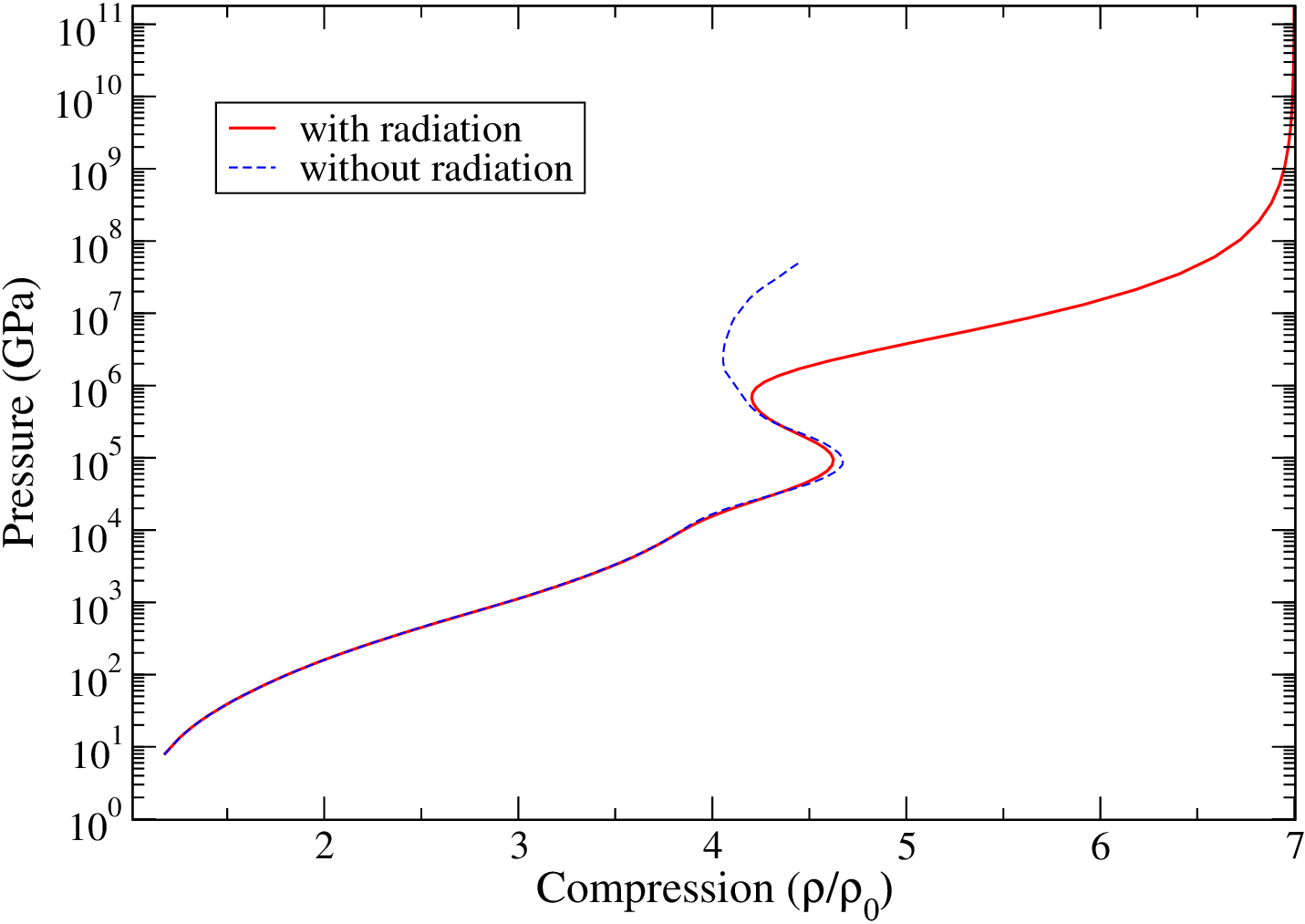}
\end{center}
\caption{Principal Hugoniot of carbon with and without blackbody radiation. }
\label{hug_C_rad}
\end{figure}

\begin{figure}[H]
\begin{center}
\vspace{1cm}
\includegraphics[width=8.5cm]{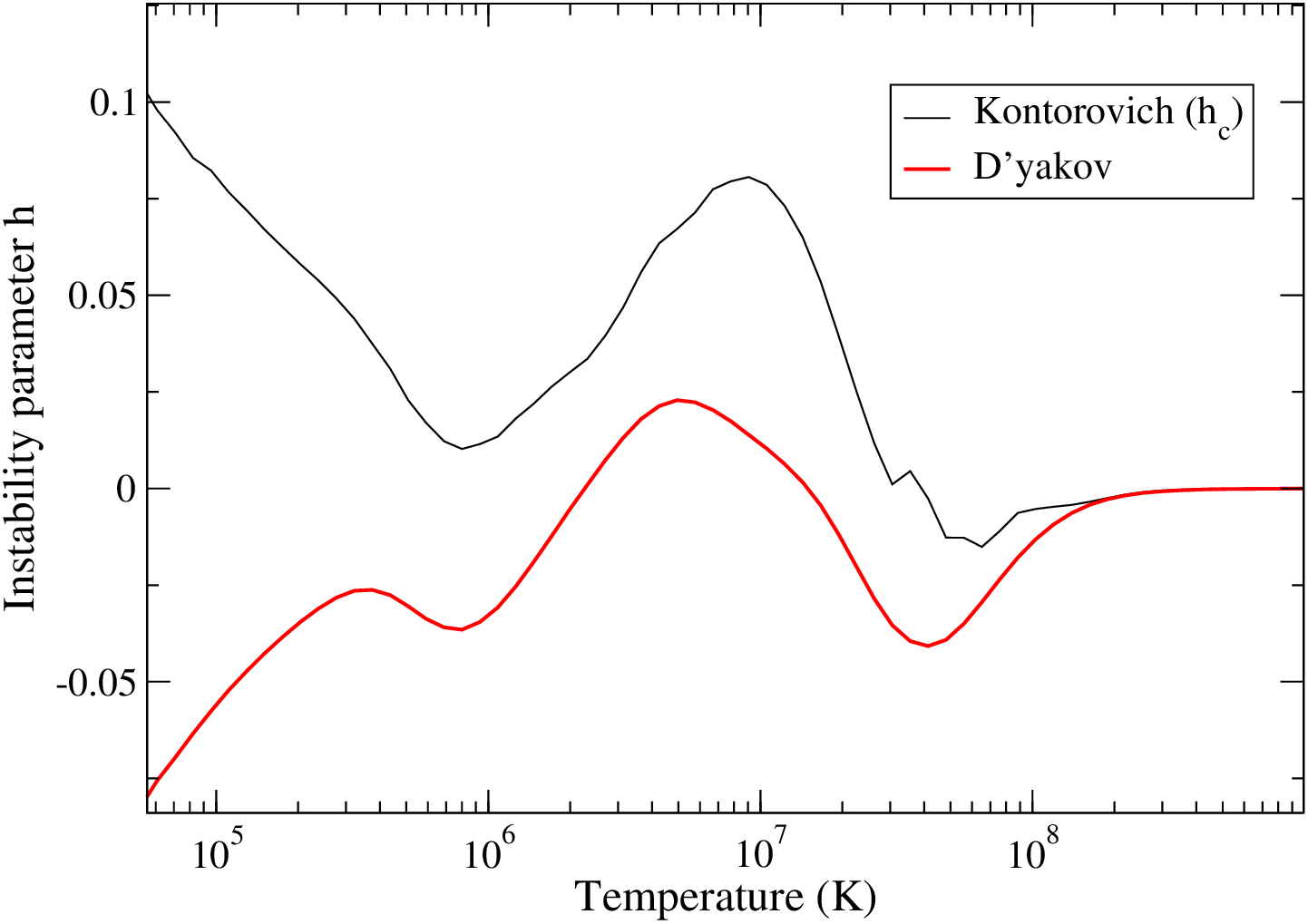}
\end{center}
\caption{Value of D'yakov's instability parameter $h$ \cite{DYAKOV54} along the principal Hugoniot of carbon, compared to the critical value $h_c$ as a function of temperature $T$. The EOS model includes blackbody radiation.}
\label{h_C_rad}
\end{figure}

\begin{figure}[H]
\begin{center}
\vspace{1cm}
\includegraphics[width=8.5cm]{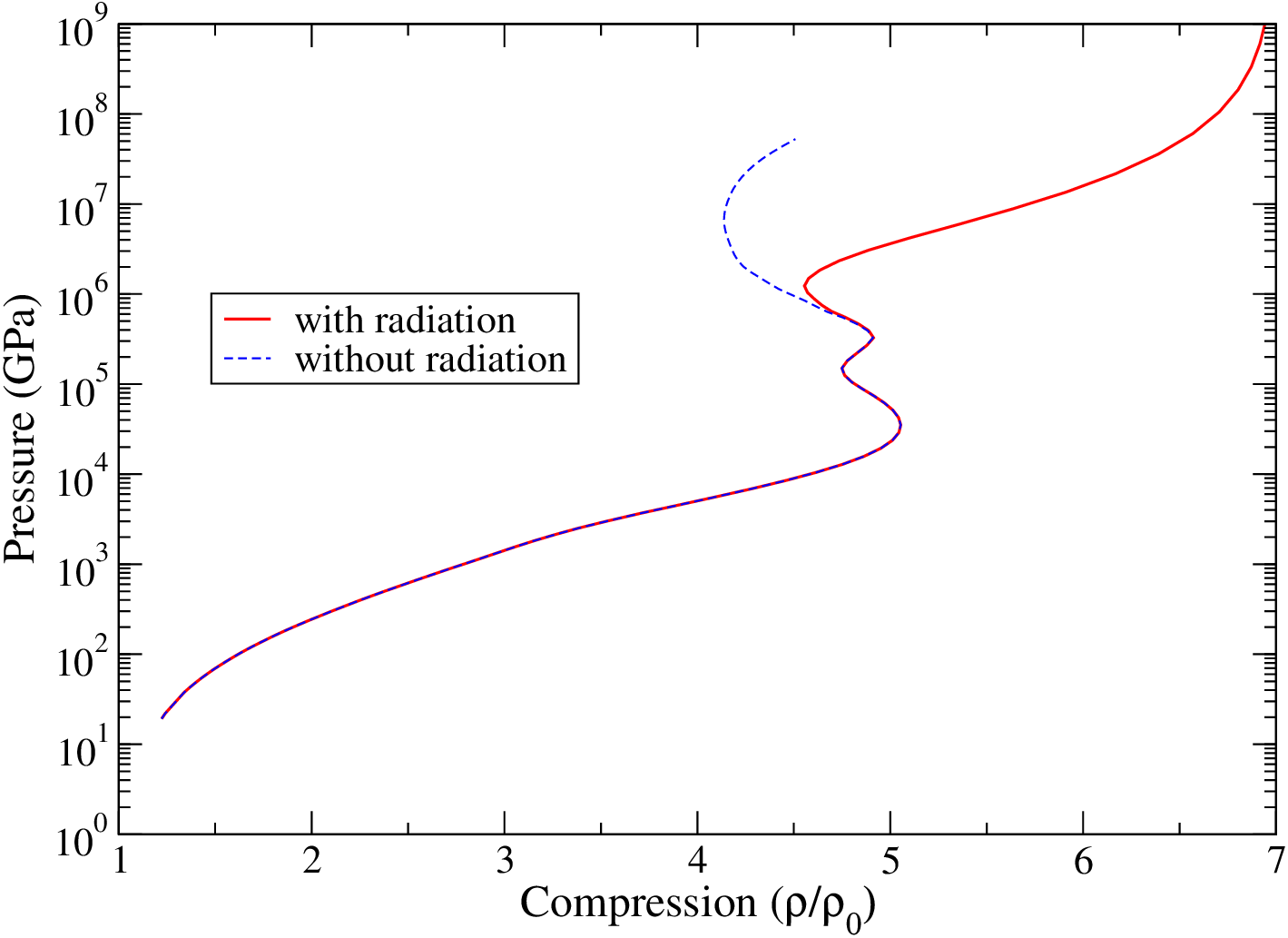}
\end{center}
\caption{Principal Hugoniot of aluminum with and without blackbody radiation.}
\label{hug_Al_rad}
\end{figure}

\begin{figure}[H]
\begin{center}
\vspace{1cm}
\includegraphics[width=8.5cm]{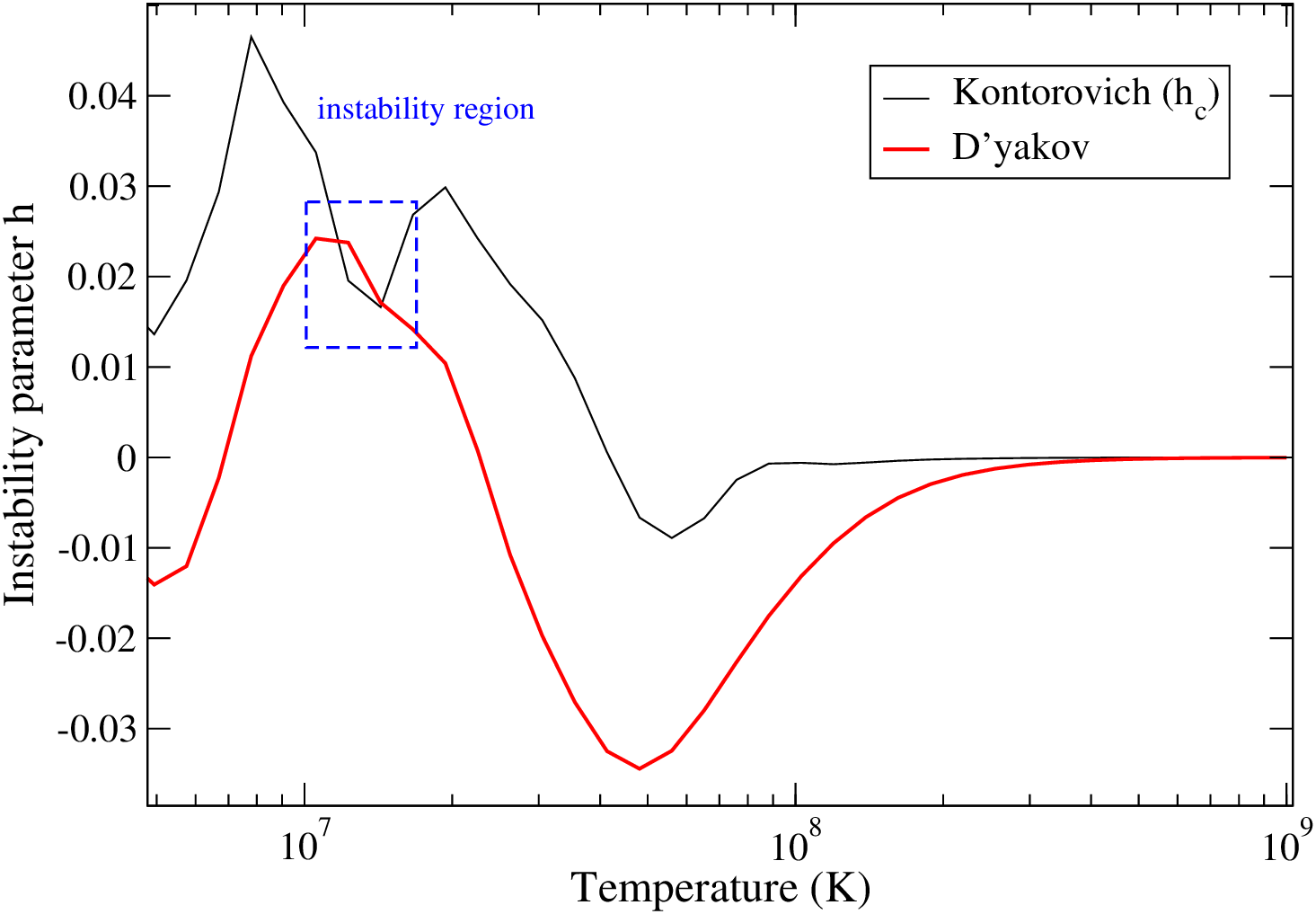}
\end{center}
\caption{Value of D'yakov's instability parameter $h$ \cite{DYAKOV54} along the principal Hugoniot of aluminum, compared to the critical value $h_c$ as a function of temperature $T$. The EOS model includes blackbody radiation.}
\label{h_Al_rad}
\end{figure}

\begin{figure}[H]
\begin{center}
\vspace{1cm}
\includegraphics[width=8.5cm]{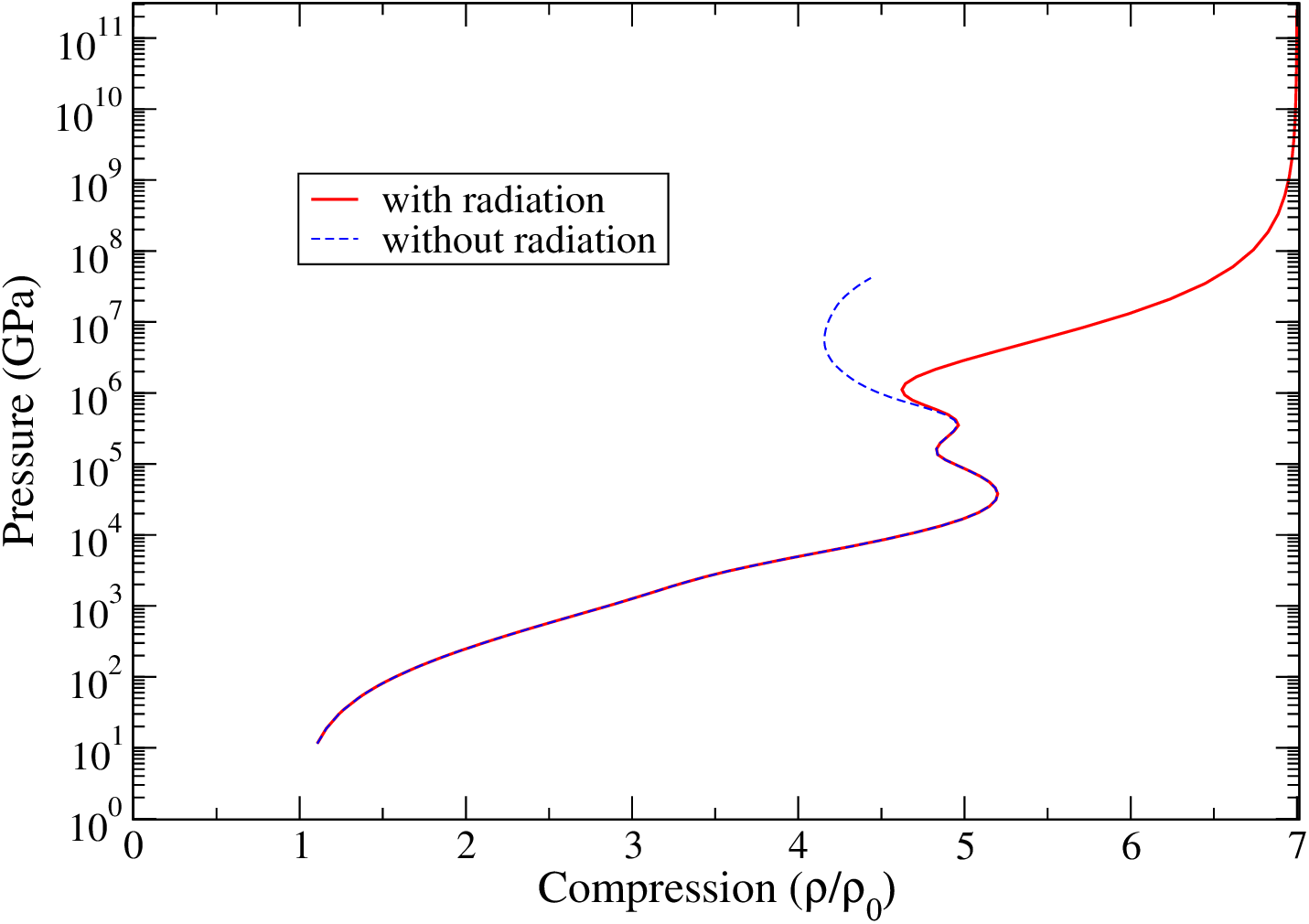}
\end{center}
\caption{Principal Hugoniot of niobium with and without blackbody radiation.}
\label{hug_Si_rad}
\end{figure}

\begin{figure}[H]
\begin{center}
\vspace{1cm}
\includegraphics[width=8.5cm]{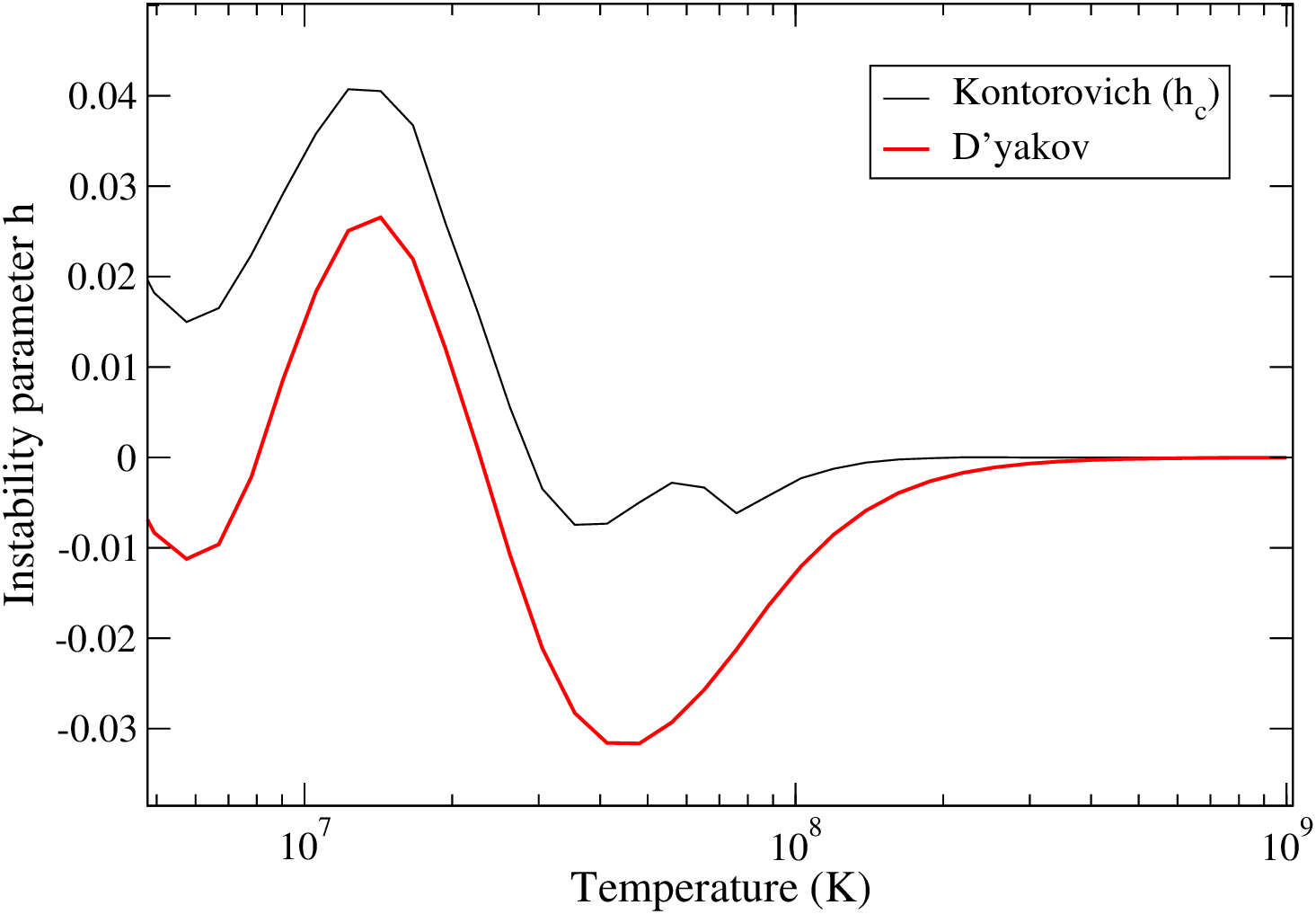}
\end{center}
\caption{Value of D'yakov's instability parameter $h$ \cite{DYAKOV54} along the principal Hugoniot of silicon, compared to the critical value $h_c$ as a function of temperature $T$. The EOS model includes blackbody radiation.}
\label{h_Si_rad}
\end{figure}

\begin{figure}[H]
\begin{center}
\vspace{1cm}
\includegraphics[width=8.5cm]{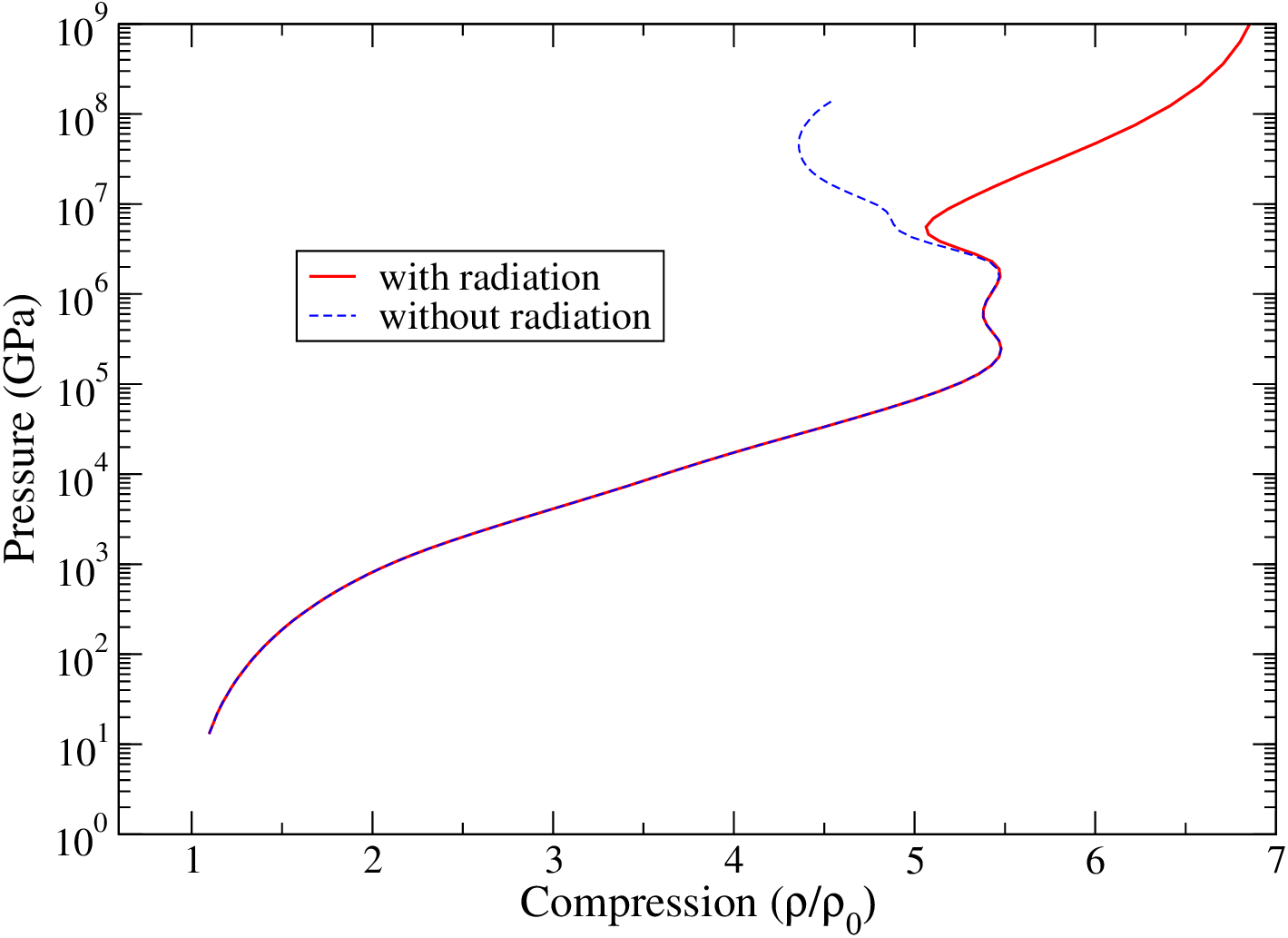}
\end{center}
\caption{Principal Hugoniot of niobium with and without blackbody radiation.}
\label{hug_Nb_rad}
\end{figure}

\begin{figure}[H]
\begin{center}
\vspace{1cm}
\includegraphics[width=8.5cm]{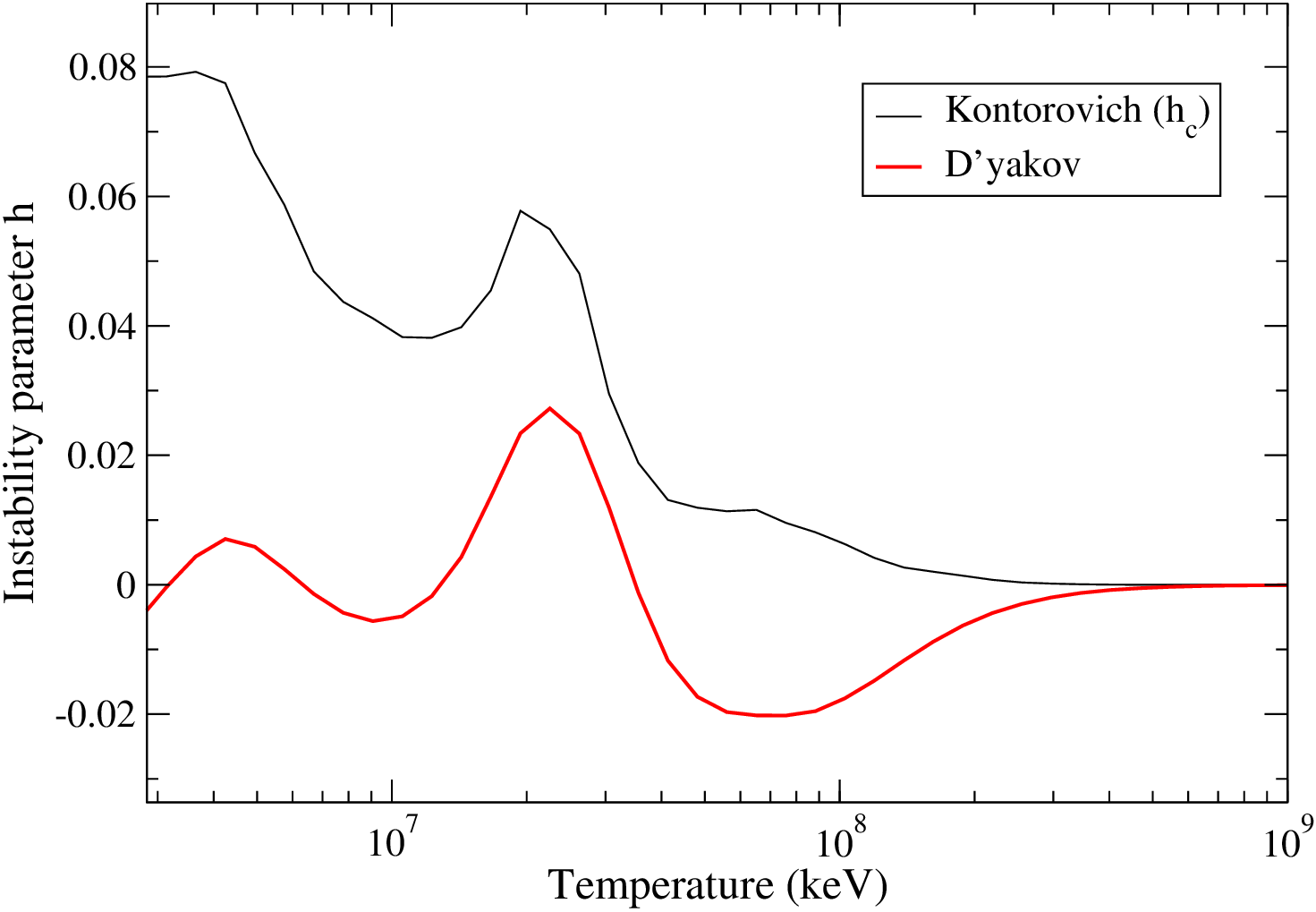}
\end{center}
\caption{Value of D'yakov's instability parameter $h$ \cite{DYAKOV54} along the principal Hugoniot of niobium, compared to the critical value $h_c$ as a function of temperature $T$. The EOS model includes blackbody radiation.}
\label{h_Nb_rad}
\end{figure}

\vspace{5mm}

The pressure and energy of equilibrium radiation can be obtained using Stefan-Boltzmann law and the energy density of radiation is given by

\begin{equation}
W=\sigma T^4,
\end{equation}

\noindent where $\sigma=\frac{\pi^2k_B^4}{60\hbar^3c^2}$=5.67$\times$10$^{-8}$ W/m$^2$/K$^4$ is Stefan-Boltsmann's constant. The free energy of the photon gas in a volume $V$ at a temperature $T$ is equal to

\begin{equation}
F_{\mathrm{rad}}=\frac{k_BTV}{\pi^2c^3}\int_0^{\infty}\omega^2\ln\left[1-e^{-\frac{\hbar\omega}{k_BT}}\right]d\omega=-PV.
\end{equation}

\noindent The entropy reads

\begin{equation}
S_{\mathrm{rad}}=-\left.\frac{\partial F_{\mathrm{rad}}}{\partial T}\right|_{V}=\frac{16\sigma}{3c}T^3V,
\end{equation}

\noindent which gives for the internal energy

\begin{equation}
E_{\mathrm{rad}}=F_{\mathrm{rad}}+TS_{\mathrm{rad}}=-3F_{\mathrm{rad}}
\end{equation}

\noindent and for the pressure

\begin{equation}\label{prad}
P_{\mathrm{rad}}=\frac{1}{3V}E_{\mathrm{rad}}=\frac{4\sigma}{3c}T^4.
\end{equation}

\noindent Identifying Eq. (\ref{prad}) with the ideal-gas law, we obtain

\begin{equation}\label{prade}
\frac{P_{\mathrm{rad}}V}{(\gamma-1)}=E_{\mathrm{rad}},
\end{equation}

\noindent with $\gamma=4/3$. The pressure $P_{\mathrm{rad}}$ (see Eq. (\ref{prad})) of the blackbody radiation at the temperature of $T$=43 keV (about 5$\times$10$^8$ K) is equal to 1.58$\times$10$^{10}$ GPa = 158 Tbar, which is orders of magnitude higher than the ``matter'' pressure (contribution of ions and electrons to the pressure, see Eq. (\ref{iocp}) and (\ref{presrel})). We can see in Figs. (\ref{hug_C_rad}), (\ref{hug_Al_rad}), (\ref{hug_Si_rad}) and (\ref{hug_Nb_rad}) that when the EOS of a photon gas is taken into account, the shock adiabat is strongly affected at high temperature, and the bifurcation to the $\rho/\rho_0$=7 asymptote starts at lower temperatures and with a higher slope than in the case where only ions and electrons are taken into account (see Eqs. (\ref{hug_C_withPIMC}), (\ref{hug_Al}), (\ref{hug_Si_new_withPIMC_new}) and (\ref{hug_Nb}) of Sec. \ref{sec3}).

When radiation is included in the EOS, the instability does not occur anymore at high temperature, which can be proven analytically (see Appendix C). There is a difference between aluminum and silicon; for silicon, the instability disappeared completely, but a narrow instability region (around $10^7$ K) remains for aluminum.

However, it is worth mentioning that the assumption of a large optical thickness, for a plasma of higher-than-solid density, is far from obvious. For instance, in the case of an aluminum plasma at a temperature of $2\times 10^7$ K and a compression close to 5, the Rosseland mean free path $\lambda_R$ is of the order of 0.45 cm. At a pressure near 10 Gbar, when the Hugoniot curves with and without radiation diverge (see Fig. \ref{hug_Al_rad}), the shock velocity is around 700 km/s. For the upper bound of the shock propagation time 1 $\mu s$, the shock travels $L$=70 cm and thus $L\gg\lambda_R$, i.e., the plasma is optically thick. On the other hand, considering the abovementioned lower bound of the shock propagation time 1 ns, the shock travels 0.07 cm and thus $L\ll\lambda_R$, i.e., the plasma is optically thin. In conditions typical from ICF experiments \cite{KRITCHER14}, the laser pulse duration is around 5 ns, and therefore the plasma is neither optically thick nor thin. Only a radiation-hydrodynamics simulation can provide reliable information about the radiation field. 

\section{Conclusion}

In this paper, we have studied the conditions for the occurrence of the D'yakov-Kontorovich hydrodynamic instability via acoustic emission in hot and dense plasmas employing an accurate EOS for electrons, based on a full quantum self-consistent relativistic (Dirac) average-atom model. The pressure is determined using the relativistic stress-tensor formula. We agree with the assertion of Das \emph{et al.} \cite{DAS11} that the instability is related to thermal as well as pressure ionization (i.e., that the shock waves become unstable for temperatures and pressures where sudden ionization of electronic shells occurs), but our conclusions are different as concerns the conditions in which the instability occurs. Our model predicts that the shock becomes unstable for higher temperatures, when the Hugoniot curve departs from the non-relativistic asymptote $\rho/\rho_0$=4 towards the limit $\rho/\rho_0=7$, i.e., almost beyond the successive ionization of the electronic shells. In other words, we find that a shock becomes unstable only for very high temperatures, at which relativistic effects on the EOS are significant. Absorption of energy from the shock wave due to ionization is a prerequisite for the occurrence of the instability, but the accounting for relativistic effects is of crucial importance. The differences between the present conclusions and the ones from Ref. \cite{DAS11} or from our previous work \cite{HEUZE09}, stem from the fact that in the models used in the two latter references, the electrons were described in a non-relativistic approximation: bound electrons were treated \emph{via} the screened hydrogenic model with $\ell$ splitting in Ref. \cite{DAS11}, and \emph{via} Schr\"odinger equation in Ref. \cite{HEUZE09}, while in both cases free electrons were modeled by the non-relativistic Thomas-Fermi model. In the present work, both bound and free electrons are described in the framework of relativistic quantum mechanics (Dirac equation in a self-consistent potential). We studied four elements: carbon, aluminum, silicon and niobium, and found that the instability occurs in a rather narrow range of thermodynamic conditions. Although we can not draw any definitive conclusion from only four examples, it seems that the instability is more likely to take place at high temperature, which is consistent with the results of Das \emph{et al.} for beryllium and aluminum, but in our case the instability occurs mostly at the end of the ionization of electronic shells, in the range where relativistic effects start to have an impact on pressure and internal energy. In such conditions, if the plasma is optically thick, the contribution of radiation should be accounted for in the energy balance. We have thus studied the impact of a blackbody radiation on the occurrence of the instability and it turns out that the Hugoniot curve departs very brutally from the compression 4 to the compression 7, preventing the instability to occur (except in a very narrow range at lower temperature for aluminum). We conclude that an accurate electronic EOS and radiative opacity is essential for a proper investigation of shock instability. Assuming that the D'yakov-Kontorovich instability exists in plasmas, its study may be a stringent test of theoretical EOS models (especially as concerns the electronic contribution), although the present works are only predictions, due to the fact that experiments in such extreme conditions remain challenging.

\vspace{1cm}

\Large

{\bf Acknowledgements}

\normalsize

\vspace{1cm}
 
The authors would like to thank M. P\'enicaud for helpful discussions about the Paradisio code and F. Soubiran for providing the DFT-MD / PIMC calculations of carbon and silicon. 

\clearpage
 
%===============================================================================
%	BIBLIOGRAPHY
%===============================================================================

\clearpage

\appendix

\section{Appendix: Non-relativistic limit of the stress-tensor formula}

Setting $V=V_{\mathrm{eff}}(R)$, $f=f(R)$ and $g=g(R)$, the derivatives $\frac{df}{dr}$ and $\frac{dg}{dr}$ at the radius of the cavity are given by

\begin{equation}
\left\{
\begin{array}{l}
\left.\frac{df}{dr}\right|_R=-\frac{\kappa}{R}f+\frac{\epsilon+c^2-V}{c}g\\
\left.\frac{dg}{dr}\right|_R=\frac{\kappa}{R}g-\frac{\epsilon-c^2-V}{c}f.
\end{array}
\right.
\end{equation}

In the non-relativistic limit: $\epsilon=k^2/2, V\ll c$, the preceding system becomes

\begin{equation}
\left\{
\begin{array}{l}
\left.\frac{df}{dr}\right|_R=-\frac{\kappa}{R}f+2cg\\
\left.\frac{dg}{dr}\right|_R=\frac{\kappa}{R}g-\frac{1}{c}\left(\frac{k^2}{2}-V\right)f
\end{array}
\right.
\end{equation}

\noindent and we deduce

\begin{equation}
\left(f\frac{dg}{dr}-g\frac{df}{dr}\right)_R=-\frac{1}{2c}\left[\left(k^2-\frac{\kappa^2}{r^2}\right)f^2+\left(\frac{df}{dr}\right)^2\right]+\frac{V}{c}f^2.
\end{equation}

Finally, noticing that

\begin{equation}
\left\{\begin{array}{l}
\sum_j(2j+1)\cdots=2\sum_{\ell}(2\ell+1)\cdots\\
\sum_j(2j+1)\kappa^2\cdots=2\sum_{\ell}(2\ell+1)(\ell^2+\ell+1)\cdots,
\end{array}
\right.
\end{equation}

\noindent we check that, assuming $V_{\mathrm{eff}}(R)=0$, Eq. (\ref{presrel}) tends to the one given by Eq. (65) of More's article \cite{MORE85}, i.e., for the free states:

\begin{equation}
P_{NR}=\frac{1}{2}\sum_{\ell}\frac{2(2\ell+1)}{4\pi R^2}\int_0^{\infty}n(\epsilon)\left[\left.\left(\frac{df}{dr}\right)\right|_R^2+\left(k^2-\frac{\ell(\ell+1)+1}{R^2}\right)f(R)^2\right]d\epsilon.\nonumber\\
\end{equation}

The non-relativistic stress-tensor expression of pressure was used in several equation-of-state models (see for instance \cite{PAIN06a,PAIN06b,PAIN07c}).

\clearpage

\section{Appendix: Proof of the asymptotic limits $\rho/\rho_0$=4 and $\rho/\rho_0$=7 for the electron gas}

In this appendix, we show that, when the radiation is not taken into account in the EOS (i.e., if we consider only ions and electrons), when the relativistic effects are negligible, the shock adiabat in the pressure-density representation tends to the limit $\rho/\rho_0$=4 and that the asymptote $\rho/\rho_0$=7 can be reached only if the electrons are ultra-relativistic.

At high temperature, $E\gg E_0$, $P\gg P_0$, and the Hugoniot relation (\ref{hug}) can be written

\begin{equation}\label{comp}
\frac{\rho}{\rho_0}=1+\frac{2E\rho}{P},
\end{equation}

and the electronic contribution to the EOS largely dominates the ionic one. Let us consider, first the non-relativistic limit, and then the relativistic one.

\begin{itemize}
\item Non-relativistic limit:
\end{itemize}

Using the non-relativistic ideal-gas equation of state: 

\begin{equation}
\left\{\begin{array}{l}
P=\rho k_BT\\
E=\frac{3}{2}k_BT,
\end{array}
\right.
\end{equation}

\noindent we get the asymptote

\begin{equation}
\frac{\rho}{\rho_0}=4.
\end{equation}

\begin{itemize}
\item Relativistic limit:
\end{itemize}

The equation of state of the relativistic ideal gas is (see for instance Ref. \cite{GREINER99}):

\begin{equation}
\left\{\begin{array}{l}
P=\rho k_BT\\
E=c^2\left\{\frac{K_1(u)}{K_2(u)}+\frac{3}{u}-1\right\}\;\;\;\; \mathrm{with} \;\;\;\; u=\frac{c^2}{k_BT}
\end{array}
\right.
\end{equation}

\noindent where $K_n(u)$ is the Bessel function of the third kind

\begin{equation}
K_n(u)=\frac{2^nu^n\Gamma(n+1/2)}{\sqrt{\pi}}\int_0^{\infty}\frac{\cos\xi}{\left(\xi^2+u^2\right)^{n+1/2}}d\xi,
\end{equation}

\noindent $\Gamma$ being the usual Gamma function \cite{ABRAMOWITZ65}. The entropy and the specific heat are respectively

\begin{equation}
S=Nk_B\left\{\ln\left[\frac{4\pi}{\rho}\left(\frac{mc}{h}\right)^3\frac{K_2(u)}{u}\right]+4+u\frac{K_1(u)}{K_2(u)}\right\}
\end{equation}

\noindent and

\begin{equation}
C_V=Nk_Bu\left\{u+\frac{3}{u}-\frac{K_1(u)}{K_2(u)}\left[3+u\frac{K_1(u)}{K_2(u)}\right]\right\}.
\end{equation}

At very high temperatures ($u\ll 1$), we can use the approximation

\begin{equation}
K_n(u)\approx\Gamma(n)\left(\frac{u}{2}\right)^{-n}
\end{equation}

\noindent where $\Gamma(n+1)=n!$. Therefore, we have

\begin{equation}\label{etkt}
E\approx c^2\frac{3}{u}=3k_BT
\end{equation}

\noindent and the asymptote in that case is, using Eq. (\ref{comp}):

\begin{equation}
\frac{\rho}{\rho_0}=7.
\end{equation}

In the present work, we do not reach conditions in which the electrons are ultra-relativistic. The strong-shock compression limit of 7 is approached when the pressure of the Planckian radiation begins to dominate. The reason is that at the temperatures much less than the rest mass of the electron 0.511 MeV, the plasma particles, rather than the photons, are responsible for the dissipations that form the shock front. However, it does not mean that the shock wave becomes relativistic or that it propagates through a photon gas. 
It is worth mentioning that the relativistic and non-relativistic limits of compression ratios of an ideal gas can be presented in a general form as \cite{ROZSNYAI14}:

\begin{equation}
\frac{P}{P_0}(\zeta)=\frac{\left[6+R(\lambda)\right]\zeta-R(\lambda)}{6+R(\lambda)-\zeta R(\lambda)}
\end{equation}

\noindent where $\zeta=\rho/\rho_0$, $\lambda=k_BT/c^2$ and the function $R(\lambda)$ is given in the Appendix of Ref. \cite{ROZSNYAI14}. We have $R(\lambda)\rightarrow 2$ as $\lambda\rightarrow 0$ and $R(\lambda)\rightarrow 1$ as $\lambda\rightarrow\infty$.

We can also evoke the fact that the strong shock compression limit for an ideal gas EOS is

\begin{equation}
\frac{\gamma+1}{\gamma-1},
\end{equation}

\noindent where $\gamma$ is the polytropic exponent. For a monoatomic gas, $\gamma=5/3$, so the compression limit is 4. For a photon gas in equilibrium, i.e., for blackbody radiation, $\gamma=4/3$, hence the limiting compression equals 7.

\clearpage

\section{Appendix: Parameters $h$ and $h_c$ in the case where radiation is included in the EOS}

The Rankine-Hugoniot relations, assuming $u_0=0$, read:

\begin{subequations}\label{hugo}
%\left\{
\begin{empheq}[left=\empheqlbrace]{align}
&VD=V_0(D-u)\label{hugo1}\\
&\left(P-P_0\right)V_0=Du\label{hugo2}\\
&\left(P+P_0\right)\left(V_0-V\right)=2\left(E-E_0\right),\label{hugo3}
\end{empheq}
%\right.
\end{subequations}

\noindent or equivalently

\begin{subequations}\label{hugo}
%\left\{
\begin{empheq}[left=\empheqlbrace]{align}
&\left(P+P_0\right)\left(V_0-V\right)=2\left(E-E_0\right)\label{hugo1a}\\
&VD=V_0(D-u)\label{hugo2a}\\
&u^2=-\left(P-P_0\right)\left(V-V_0\right),\label{hugo3a}
\end{empheq}
%\right.
\end{subequations}

\noindent where the EOS of the pole $\left(P_0,E_0\right)$ is dominated by ions and electrons, and where the temperature is sufficiently high so that the point $\left(P,E\right)$ on the shock adiabat is dominated by the radiation field.

We have, using Eq. (\ref{hugo2a}) and (\ref{hugo3a}):

\begin{equation}
(D-u)^2=\frac{\left(P-P_0\right)V}{\zeta-1},
\end{equation}

\noindent where $\zeta=V_0/V=\rho/\rho_0$ is the compression. Taking $E=E_{\mathrm{rad}}$ (see Eq. (\ref{prade}) and \ref{etkt}) and differentiating Eq. (\ref{hugo1a}) yields

\begin{equation}
dP\left(V_0-V\right)-\left(P+P_0\right)dV=6PdV+6VdP
\end{equation}

\noindent and since

\begin{equation}
\left.\left.\frac{d\rho}{dP}\right|_{\mathcal{H}}=-\frac{1}{V^2}\frac{dV}{dP}\right|_{\mathcal{H}},
\end{equation}

\noindent we obtain

\begin{equation}
\left.\frac{d\rho}{d P}\right|_{\mathcal{H}}=\frac{7-\zeta}{\left(P_0+7P\right)V}.
\end{equation}

\noindent The D'yakov instability parameter

\begin{equation}
h=\left.-(D-u)^2\frac{d\rho}{dP}\right|_{\mathcal{H}},
\end{equation}

\noindent is therefore equal to

\begin{equation}
h=\frac{\left(\frac{P}{P_0}-1\right)}{\left(1+7\frac{P}{P_0}\right)}\frac{(7-\zeta)}{(1-\zeta)}.
\end{equation}

\noindent Combining Eqs. (\ref{prade}) and (\ref{hugo1a}) gives

%\begin{equation}
%\frac{P}{P_0}=\frac{(1-7\zeta)}{(\zeta-7)}
%\end{equation}

\begin{equation}
\frac{P}{P_0}=\frac{1-(1+2\chi_0)\zeta}{\zeta-7}
\end{equation}

\noindent where 

\begin{equation}
\chi_0=\frac{E_0}{P_0V_0}
\end{equation}

\noindent and finally

\begin{equation}\label{trueh}
h=\frac{(\zeta-7)\left[\left(1+\chi_0\right)\zeta-4\right]}{\zeta(\zeta-1)\left(3+7\chi_0\right)}
\end{equation}

\noindent which reduces, in the case of a pure photon gas ($E_0=3P_0V_0$) to

\begin{equation}
h=\frac{\zeta-7}{6\zeta}.
\end{equation}

\begin{table}
\begin{center}
\begin{tabular}{cc}\hline\hline
Element & Value of $\chi_0$ \\\hline
C & 1.85\\
Al & 1.99\\
Si & 1.95\\
Nb & 2.02\\\hline\hline
\end{tabular}
\caption{Values of $\chi_0=\frac{E_0}{P_0V_0}$ for the different elements considered in the present paper.} \label{tabchi0}
\end{center}
\end{table}

\noindent The square of the sound speed can be expressed as

\begin{equation}
c_s^2=\left.\frac{\partial P}{\partial\rho}\right|_T+\frac{T}{\rho^2}\frac{\left(\left.\frac{\partial P}{\partial T}\right|_{\rho}\right)^2}{\left.\frac{\partial E}{\partial T}\right|_{\rho}}.
\end{equation}

\noindent With $E=E_{\mathrm{rad}}=\frac{4\sigma}{c}T^4V$ and $P=P_{\mathrm{rad}}=\frac{4\sigma}{3c}T^4$, we have

\begin{equation}
\left.\frac{\partial P}{\partial\rho}\right|_T=0
\end{equation}

\noindent and

\begin{equation}
c_s^2=\frac{4}{3}PV,
\end{equation}

\noindent which is a particular case of $c_s^2=\gamma PV$ for $\gamma=4/3$. The Mach number becomes

\begin{equation}
M^2=\frac{(D-u)^2}{c_s^2}=\frac{3}{2(\zeta-1)}\frac{\left[\left(1+\chi_0\right)\zeta-4\right]}{\left[\left(1+2\chi_0\right)\zeta-1\right]}
\end{equation}

\noindent and reduces, in the case of a pure photon gas ($E_0=3P_0V_0$), to

\begin{equation}
M^2=\frac{6}{7\zeta-1}
\end{equation}

\noindent and the critical parameter is

\begin{equation}\label{truehc}
h_c=\frac{1-(\zeta+1)M^2}{1-(\zeta-1)M^2}=\frac{(\zeta-7)\left[\zeta\left(\chi_0-1\right)-2\right]}{(\zeta-1)\left[\zeta\left(5+7\chi_0\right)-14\right]}
\end{equation}

\noindent which reduces, in the case of a pure photon gas, to

\begin{equation}
h_c=\frac{\zeta-7}{13\zeta-7}.
\end{equation}

\noindent Therefore $h>h_c$ implies that $\zeta<\zeta_t$, where $\zeta_t$ is a positive real solution of the second-order equation

\begin{equation}\label{zetat}
2\left(1+2\chi_0\right)\zeta^2-7\left(1+\chi_0\right)\zeta+14=0.
\end{equation}

\begin{figure}[H]
\begin{center}
\vspace{1cm}
\includegraphics[width=8.5cm]{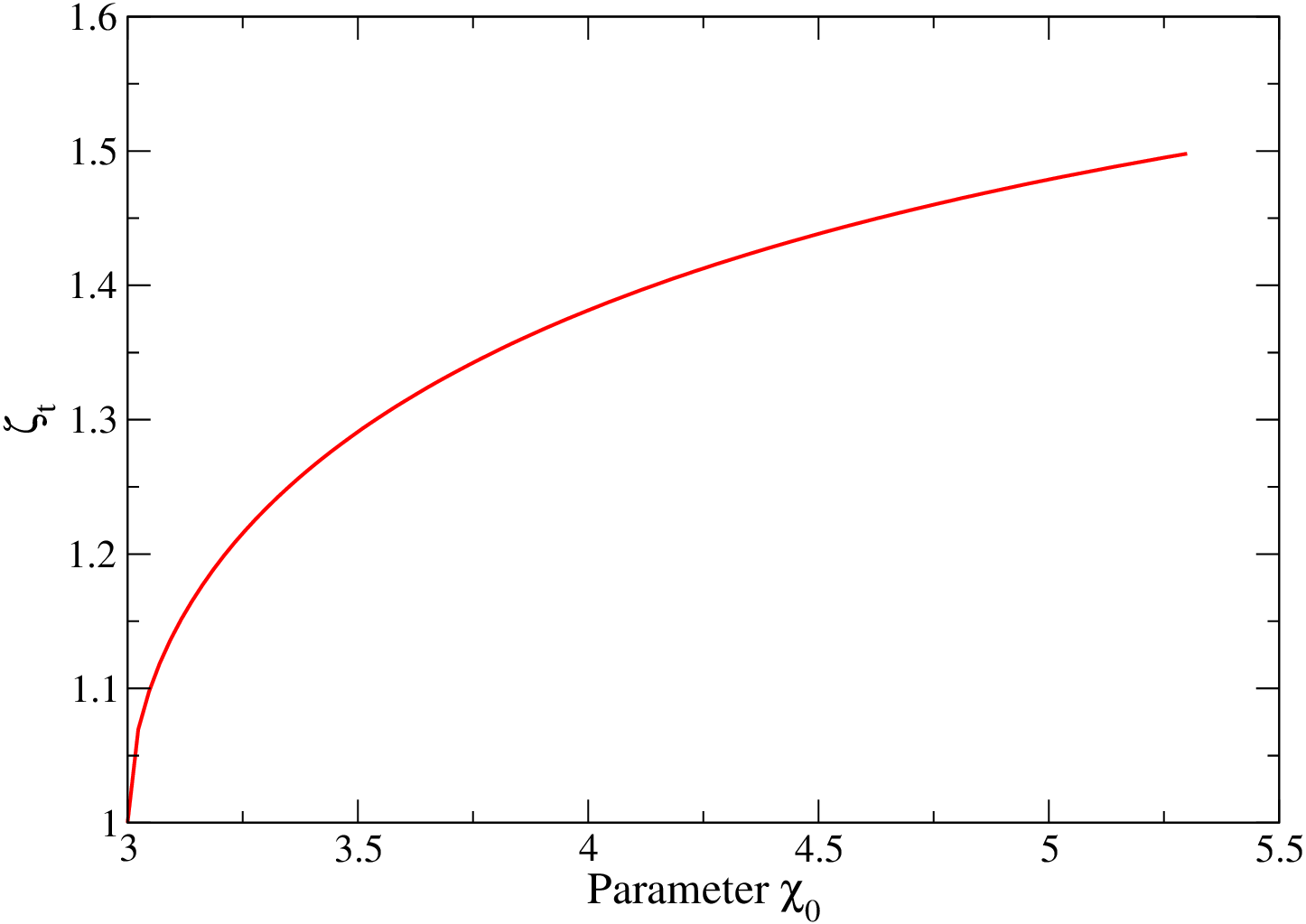}
\end{center}
\caption{Value of $\zeta_t$ as a function of parameter $\chi_0$.}
\label{fig26}
\end{figure}

\begin{figure}[H]
\begin{center}
\vspace{1cm}
\includegraphics[width=8.5cm]{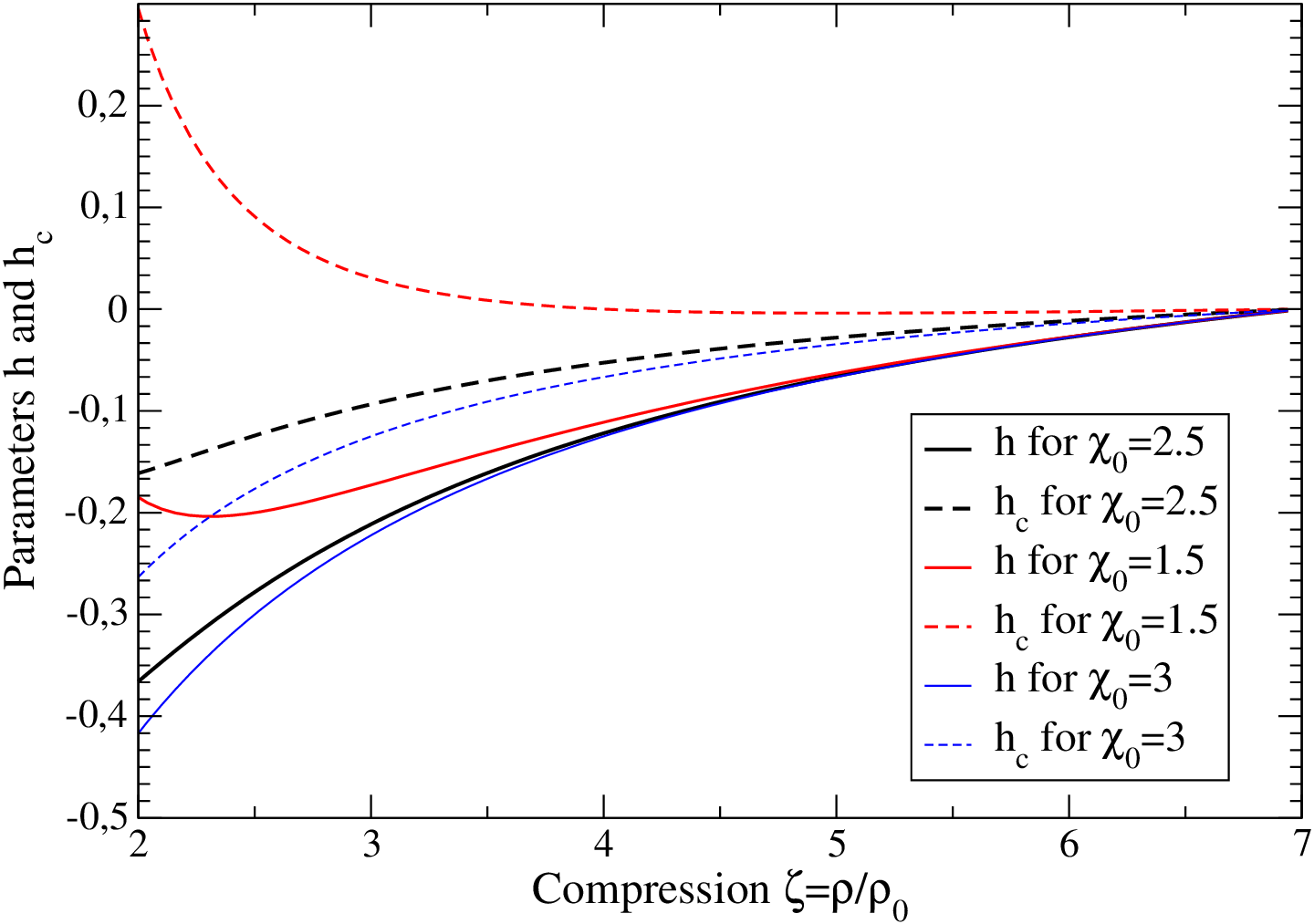}
\end{center}
\caption{Variation of parameters $h$ and $h_c$ versus compression $\zeta=\rho/\rho_0$ for three different values of parameter $\chi_0$: 1.5, 2.5 and 3. The conditions in which the photon gas is the dominant contribution to the EOS in the present work start at $\zeta\approx$ 5.}
\label{fig27}
\end{figure}

As can be seen in figure \ref{fig26}, in the range $0<\chi_0<3$, which includes the values of our EOS model (see table \ref{tabchi0}), Eq. (\ref{zetat}) has no real solution (for $\chi_0$=3, the only real solution is $\zeta=1$). Therefore, we always have $h<h_c$, which means that when the radiation dominates, the system is always stable (see Fig. \ref{fig27}).

Of course, the case $\chi_0$=3, corresponding to a pure photon gas, is not physical. Naturally, shock waves cannot exist in a photon gas because a shock wave is a non-linear flow. Non-linear hydrodynamic equations used by D'yakov and Kontorovich can describe shock fronts. To consider shock fronts in a photon gas, one needs a significant photon-photon scattering, a non-linear QED effect, which is only possible if the corresponding electromagnetic field strength exceeds the Schwinger limit \cite{BUCHANAN06,SCHWINGER51,BULANOV10}, allowing the interacting photons to produce virtual electron-positron pairs, which means that the temperature of the photon gas must be in MeV range and that the post-shock pressure should exceed $\approx 10^{10}$ Gbar, i.e., $10^{15}$ GPa, orders of magnitude above the pressure range studied in the present work.

\end{document}